\documentclass[12pt]{article}
\usepackage{longtable}
\usepackage[section]{placeins}
\usepackage[top=1in, left=1in, right=1in, bottom=1in]{geometry}
\usepackage[parfill]{parskip}	
\usepackage{graphicx}
\usepackage{caption}
\usepackage{enumerate}
\usepackage{subcaption}
\usepackage[export]{adjustbox}
\usepackage{pdfpages}
\usepackage{amssymb}
\usepackage{xcolor}
\usepackage{multirow}
\usepackage{changepage}
\usepackage{epstopdf}
\usepackage{bbm}
\usepackage{bm}
\usepackage{multicol}
\usepackage{amssymb}
\usepackage{amsmath}
\usepackage{amsfonts}
\DeclareMathAlphabet{\mathbbmsl}{U}{bbm}{m}{sl}

\usepackage{lscape}
\usepackage{rotating}
\usepackage{setspace}
\usepackage{threeparttable}
\usepackage{booktabs}
\usepackage[compact]{titlesec}
\usepackage{lmodern}
\fontfamily{lmtt}\selectfont
\usepackage[T1]{fontenc}

\usepackage[comma,sort&compress]{natbib}
\usepackage{hyperref}
\usepackage{titling}
\usepackage{pifont}
\usepackage{soul}
\usepackage[capposition=top]{floatrow}
\newcommand{\subtitle}[1]{%
  \posttitle{%
    \par\end{center}
    \begin{center}\large#1\end{center}
    \vskip0.5em}%
}
\usepackage{amsthm}
\usepackage{amsmath} 
\usepackage{multirow}

\usepackage{listings}

\usepackage{mathrsfs}

\usepackage[edges]{forest}
\usepackage{subcaption}

\newtheorem*{remark*}{Remark}

\usepackage{xpatch}
\makeatletter
\xpatchcmd{\@thm}{\thm@headpunct{.}}{\thm@headpunct{}}{}{}
\makeatother

\usepackage{mathrsfs}

\usepackage{color}
\begin{document}
\pagestyle{plain}

\newtheoremstyle{mystyle}
{\topsep}
{\topsep}
{\it}
{}
{\bf}
{.}
{.5em}
{}
\theoremstyle{mystyle}
\newtheorem{assumptionex}{Assumption}
\newenvironment{assumption}
  {\pushQED{\qed}\renewcommand{\qedsymbol}{}\assumptionex}
  {\popQED\endassumptionex}
\newtheorem{assumptionexp}{Assumption}
\newenvironment{assumptionp}
  {\pushQED{\qed}\renewcommand{\qedsymbol}{}\assumptionexp}
  {\popQED\endassumptionexp}
\renewcommand{\theassumptionexp}{\arabic{assumptionexp}$'$}

\newtheorem{assumptionexpp}{Assumption}
\newenvironment{assumptionpp}
  {\pushQED{\qed}\renewcommand{\qedsymbol}{}\assumptionexpp}
  {\popQED\endassumptionexpp}
\renewcommand{\theassumptionexpp}{\arabic{assumptionexpp}$''$}

\newtheorem{assumptionexppp}{Assumption}
\newenvironment{assumptionppp}
  {\pushQED{\qed}\renewcommand{\qedsymbol}{}\assumptionexppp}
  {\popQED\endassumptionexppp}
\renewcommand{\theassumptionexppp}{\arabic{assumptionexppp}$'''$}

\renewcommand{\arraystretch}{1.3}

\newcommand{\argmin}{\mathop{\mathrm{argmin}}}
\makeatletter
\newcommand{\grande}{\bBigg@{2.25}}
\newcommand{\enorme}{\bBigg@{5}}

\newcommand{\blind}{0}

\newcommand{\tit}{\Large Risk Set Matched Difference-in-Differences for the Analysis of Effect Modification in an Observational Study on the \\ Impact of Gun Violence on Health Outcomes}

\author{Eric R. Cohn\thanks{Department of Biostatistics, Harvard School of Public Health, 677 Huntington Avenue, Boston, MA 02115; email: \url{ericcohn@g.harvard.edu}.} \and Zirui Song\thanks{Department of Health Care Policy, Harvard University and Department of Medicine, Massachusetts General Hospital, 180 A Longwood Avenue
Boston, MA 02115; email: \url{song@hcp.med.harvard.edu}.} \and Jos\'{e} R. Zubizarreta\thanks{Department of Health Care Policy, Biostatistics, and Statistics, Harvard University, 180 A Longwood Avenue, Office 307-D, Boston, MA 02115; email: \url{zubizarreta@hcp.med.harvard.edu}}}

\date{} 

\if0\blind
\title{ \tit}
\date{} 
\maketitle
\fi
\vspace{-40pt}

\begin{abstract}
\textcolor{black}{Gun violence is a major source of injury and death in the United States.}
However, relatively little is known about the effects of firearm injuries on survivors and their family members and how these effects vary.
To study these questions and, more generally, to address a gap in the methodological causal inference literature, we present a framework for the study of effect modification or heterogeneous treatment effects in difference-in-differences designs.
We implement a new matching technique, combining profile matching and risk set matching, to (i) preserve the time alignment of covariates, exposure, and outcomes, avoiding pitfalls of other common approaches for difference-in-differences, and (ii) explicitly control biases due to imbalances in observed covariates in subgroups discovered from the data.
Our case study shows significant and persistent effects of nonfatal firearm injuries on several health outcomes for those injured and on the mental health of their family members.
Sensitivity analyses reveal that these results are moderately robust to unmeasured confounding bias.
Finally, while the effects for those injured \textcolor{black}{vary} largely by the severity of the injury and its documented intent, for families, effects are strongest for those whose relative's injury is documented as resulting from an assault, self-harm, or law enforcement intervention.
\end{abstract}

\begin{center}
\end{center}

\newpage
\doublespacing
\section{Introduction}
\label{sec_intro}

\subsection{Impact of gun violence on those injured and their families}
\label{subsec_gunviolence}
Firearms are a leading cause of morbidity and mortality in the United States.
\textcolor{black}{The economic costs to society are large as well, with recent estimates placing the total cost of firearm injuries at \$557 billion annually \citep{SongZirui2022TBCf}.}
\textcolor{black}{Accordingly}, health and social sciences researchers have \textcolor{black}{increasingly} pursued an understanding of the wide-ranging implications of exposure to firearm violence \citep{SubbaramanNidhi2021Gvis}.

In particular, associations between firearm violence and poverty \citep{10.1001/jamainternmed.2022.5460}, trauma \citep{MageeLaurenA.2022Mhof}, and racism \citep{FitzpatrickVeronica2019NFVT} have all been documented, with variation by type of exposure (e.g., suicide versus homicide) and other characteristics (e.g., age \textcolor{black}{of the victim}); see also \cite{FowlerKatherineA2015Fiit}, \cite{WintemuteGarenJ2015TEoF}, \cite{KaufmanElinoreJ2021ETiF}, and \cite{schumacher_kirzinger_presiado_valdes_brodie_2023}.

However, many of these studies do not rigorously address the difficulties of making causal inferences from observational data.
Specifically, they do not clearly state causal estimands or identification assumptions, carefully control for confounding, or implement sensitivity analyses to assess the robustness of findings to violations of key assumptions.
Additionally, many limit their study to firearm deaths rather than nonfatal injuries, which affect larger portions of the population (roughly, two to three times more in the United States; \citealt{KalesanBindu2017THEo, KaufmanElinoreJ2021ETiF}), and do not assess the impacts on family members, \textcolor{black}{who are also meaningfully} affected.
Furthermore, estimates of economic effects have often lacked the data to capture the full scope of societal spending\textcolor{black}{, often relying on hospital discharge data that contain hospital charges (or list prices) rather than actual transacted prices paid by insurers and patients.}

To address these gaps, we extend an analysis by \cite{SongZirui2022CiHC} by examining a large longitudinal data set of health care claims and estimate the heterogeneous causal effects of nonfatal firearm injuries on the health and health care of those injured and their families.
Among others, we use a more comprehensive dataset and present methods for the discovery of effect modification in difference-in-differences studies.
Our study results can help clinicians and policymakers understand and anticipate the profound clinical and economic effects of firearm injuries, contributing to a growing focus on evidence-based prevention efforts.

\subsection{Effect modification in difference-in-differences studies}
\label{subsec_DiDHTE}

\subsubsection{Review of methods for difference-in-differences}
\label{subsubsec_DiDrev}
Our observational study entails the difference-in-differences design, a widely used approach for causal inference in observational studies with longitudinal data.
Originating in the seminal studies of \cite{SnowOMCC} and \cite{CardD1994Mwae}, the canonical setting involves outcome data from two time periods (one pre- and one post-exposure) and an assumption that the exposed group's outcomes would evolve in parallel to the control group's in the absence of exposure; that is, the assumption of parallel trends or additive equiconfounding \citep{SoferTamar2016ONOC}.
The method has since been extended to several other settings, including, for example, to those where parallel trends hold conditionally on covariates \citep{StuartElizabethA2014Upsi} or only approximately \citep{KeeleHasegawa2019}, to those where units are exposed in a staggered manner at different points in time \citep{Ben‐MichaelEli2022Scws}, and to non-linear outcomes \citep{AtheySusan2006IaIi} and instrumental variables  \citep{ye2022instrumented}.
See \cite{ding2019bracketing} and \cite{RothJonathan2022WTiD} for perspectives and reviews on difference-in-differences designs.

Estimation in difference-in-differences often involves fitting a linear regression model, where the outcomes are regressed on an indicator of being in the post-exposure period and the exposed group, along with two-way fixed effects (TWFE) for each unit and each time.
In settings where the exposure is staggered, however, traditional TWFE can be problematic \citep{BakerAndrewC.2022Hmsw}.
\textcolor{black}{In such settings, treatment effect estimates comprise weighted averages of estimates from each time period. As discussed by \cite{Goodman-BaconAndrew2021Dwvi}, this approach can give rise to complications.}
\textcolor{black}{First, i}n the presence of certain forms of effect heterogeneity, \textcolor{black}{the weights can be negative and the overall estimate} can involve contaminated comparisons, where earlier-exposed units are used as controls for later-exposed units.
\textcolor{black}{Second, even in the absence of such comparisons, the weighted average may not correspond to a policy-relevant estimand without additional assumptions, and each time point can receive a non-intuitive weight.}
More advanced TWFE methods have been developed to avoid this issue \citep{BorusyakRESD, CallawayBrantly2021Dwmt, SunLiyang2021Edte}.
However, as with any parametric regression method, model misspecification remains a concern.

Matching \citep{StuartElizabethA2014Upsi, basu2020constructing, ImaiKosuke2021MMfC, BlackwellMatthew2022Tmfr} is a covariate adjustment method that explicitly controls which units' outcomes are compared, reduces the dependence on parametric models, and is straightforward to implement and interpret.
One such method for longitudinal data is risk set matching \citep{LiYunfeiPaul2001BRSM}, which, at each time, matches a newly exposed unit with controls that are not yet exposed based on past covariate information. 
Once a control is used in a match, it is discarded from the pool of potential controls for later-exposed units.
Risk set matching thus entails tailored comparisons for each exposed unit, where that unit's outcomes are compared only to those of units not yet exposed, and these units are not recycled.
Further, the resulting matched data structure enables the use of familiar methods for randomization inference and sensitivity analysis, which are key components of our analytical framework and observational study.

\subsubsection{Review of methods for effect modification}
\label{subsubsec_HTErev}
Our study seeks not only to estimate the main effects of nonfatal firearm injury, but also to understand how these effects vary across individuals.
Several methods exist to test for effect modification by observed covariates (e.g., \citealt{DingPeng2016Rift}), and it is common for studies to include subgroup analyses to estimate effects within pre-specified subgroups of policy or practical relevance.
Machine learning prediction methods have also been extended to causal inference settings in order to discover effect modification across complex and possibly unanticipated interactions of observed covariates \citep{HillJenniferL.2011BNMf, HsuJesseY.2015Scot, AtheySusan2016Rpfh,  https://doi.org/10.48550/arxiv.2009.09036, YangJiabei2022CitF}.
Finally, effect modification relates to the design sensitivity of an observational study, as estimates for subgroups with larger effects may be less sensitive to unmeasured confounding bias \citep{HsuJesseY2013EMaD}.

Two recent methods, the submax \citep{LeeKwonsang2018Apat} and de novo \citep{LeeKwonsang2021DHEE} methods, test for the presence of effect modification in matched observational studies by both pre-specified effect modifiers and those discovered from the data.
Both compute subgroup-specific test statistics and use their joint randomization distribution for inference.
While the former uses pre-specified subgroups, the latter uses data splitting to discover promising subgroups in the first half of the data and assesses statistical significance in the second half.
A particular advantage of these methods is that they allow the use of familiar sensitivity analyses to assess the robustness of results to unmeasured confounding bias \citep{RosenbaumPaulR2020DoOS}.
However, the submax method, the denovo method, and many others for effect modification are developed under an assumption of strong ignorability rather than parallel trends.

In difference-in-differences, approaches for testing for effect modification often entail so-called difference-in-difference-in-differences (DiDiD or DDD; \citealt{LongSharonK2010DtEo, FredrikssonAnders2019IeuD}) comparisons and involve computing difference-in-differences estimates for multiple pre-specified subgroups.
Methods for inference are often parametric and assess the statistical significance of interaction terms in TWFE regressions (e.g., \citealt{YuWenhua2020Tabl, LeifheitKathrynM.2021EEMa}) or compute subgroup-specific asymptotic confidence intervals (e.g., \citealt{ChartersThomasJ.2013Eoao}).
Given the limitations of TWFE expressed above, these approaches may be problematic.
For example, like their main term counterparts, coefficient estimates on interaction terms may lack their intended interpretation as weighted averages of treatment effects in staggered settings, as in our study \citep{RothJonathan2022WTiD}.

\subsection{Covariate balance in the study of effect modification}
Most methods for understanding effect modification are agnostic to covariate balance, a key ingredient in the design of both experimental and observational studies.
Methods can then perform poorly in finite samples when subgroups are imbalanced.
For example, \cite{RigdonJoseph2018Pfdo} show how imbalances in discovered subgroups can inflate Type I errors in randomized experiments that use regression with variable selection or random forests to discover effect modifiers.
Popular covariate adjustment methods, such as regression or matching, typically control bias from observed covariates by balancing covariates within only a few pre-specified subgroups (i.e., by including subgroup indicators in the model or matching within strata, respectively).
Such approaches are untenable when subgroups are discovered rather than pre-specified, as in our study.

\subsection{Contribution and outline}
\label{subsec_outline}
In this paper, we estimate the causal impacts of nonfatal firearm injuries on those injured and their family members and the \textcolor{black}{variation in} these effects by \textcolor{black}{levels of the exposure and} observed individual and community characteristics.
To accomplish this, we present a randomization-based inferential framework \citep{AtheySusan2022DaiD, FogartyColinB.2022Twni} for the study of effect modification \textcolor{black}{and} heterogeneous treatment effects in difference-in-differences settings with staggered exposures, which to our knowledge has not been formalized in the causal inference literature.
Following \cite{RosenbaumPaulR2020DoOS}, this framework delineates assumptions for identification, mode of inference, and sensitivity analysis for such difference-in-differences designs.
\textcolor{black}{In particular, our sensitivity analysis extends work by \cite{Keele2019} to handle longitudinal data, matched sets of varying size, and staggered treatment adoption.}

In addition, we describe a new matching technique which combines profile matching \citep{CohnEricR2022PMft} with risk set matching \citep{LiYunfeiPaul2001BRSM} to respect the time alignment of the exposure, outcomes, and covariates and to strongly control bias due to observed covariate imbalances in subgroups discovered from the data.
Finally, we make two more technical contributions by extending (i) the submax-method of \cite{LeeKwonsang2018Apat}, which tests for stronger design sensitivity across pre-specified subgroups, to instead test for the presence of effect variation \textcolor{black}{for groups defined by levels of the exposure} and (ii) the de novo method of \cite{LeeKwonsang2021DHEE}, which discovers effect modifiers using recursive partitioning, from matched pairs to matched sets of varying size.

The rest of this paper is organized as follows.
Section \ref{sec_data} describes a large administrative data set of health care claims used to study the impacts of nonfatal firearm injury on those injured and their families.
Section \ref{sec_framework} introduces the mathematical notation, matching method, \textcolor{black}{sensitivity analysis model,} and statistical approach used to obtain causal estimates from the  administrative data set.
Section \ref{sec_impacts} evaluates the quality of the matched design, estimates average effects of nonfatal firearm injury for a variety of outcomes, and explores \textcolor{black}{variation} of these effects across both pre-specified \textcolor{black}{levels of the exposure} and subgroups discovered from the data.
In the Supplementary Materials, we evaluate the extension of the de novo method in a simulation study.
Section \ref{sec_conclusion} concludes.

\section{A large data set on health care claims}
\label{sec_data}
The data used for analysis comprise health care claims and enrollment data from one of the United States's largest databases, MarketScan, with data collected between 2008 and \textcolor{black}{2021}.
This database includes \textcolor{black}{detailed} claims and enrollment information for more than 40 million individuals each year, and it provides a unique opportunity to follow individuals over time to track the evolution of a variety of \textcolor{black}{health} outcomes.
The commercially insured enrollees had employer-sponsored health insurance, generally from large U.S. employers, and the Medicare enrollees were Medicare beneficiaries with Medicare Supplemental (i.e., ``Medigap'') coverage.

\color{black}
These data are particularly well-suited for our primary and secondary substantive research objectives, which are, respectively, (i) to estimate the causal effects of nonfatal firearm injuries on the health care outcomes of survivors and their families and (ii) to discover and estimate heterogeneity in these effects.
As described in detail in \cite{SongZirui2022CiHC}, the sheer number of individuals included in our databases contributes to their representativeness, and the detailed data on patient claims allows estimation of \textcolor{black}{actual transacted health care spending} absent in many prior studies.
Additionally, data in prior studies have lacked information on mental and behavioral health conditions, which may be particularly important to understand for family members of those injured by firearms.
The individual-level data also allow us to track patients over time to estimate long-run effects.
A potential limitation of our data is the omission of Medicaid or uninsured patients, which may limit the generalizability of results to these populations.

\color{black}
We assess the causal effects of two individual-level exposures, (1) experiencing a non-fatal firearm injury and (2) having a family member with a non-fatal firearm injury.
We limit our analysis to individuals continuously enrolled at least one year before and one year after the injury.
International Classification of Diseases, Ninth Revision (ICD-9) and Tenth Revision (ICD-10) codes were used to identify firearm injuries in the claims data.
Injured people with and without family members were included, and family members were identified using family relationships in the data. 
Covariates available in the data include age, Diagnostics Cost Groups (DxCG) risk score, binary sex, health insurance plan type, and whether the plan covers prescription drugs.
The DxCG risk score is a measure of health status or expected spending that is commonly used for risk adjustment, and it is calculated annually for each individual using demographic and clinical diagnosis data.

Following a more limited analysis in \cite{SongZirui2022CiHC}, we assess the effects of nonfatal firearm injury on seven outcomes: (1) medical spending (adjusted for inflation to 2019 U.S. dollars); receipt of (2) pain, (3) psychiatric, or (4) substance use disorder diagnoses; and the number of days of (5) pain, (6) psychiatric, or (7) other medications prescribed and dispensed among those with prescription drug coverage.
Our longitudinal data afford the opportunity to examine effects on multiple timescales.
That is, we estimate the change in each outcome compared to the previous twelve months with respect to (1) the month immediately after the injury and (2) the average over the twelve months post-injury. 


\section{Effect modification in difference-in-differences studies}
\label{sec_framework}

\subsection{Setup, notation, and \textcolor{black}{basic} assumptions}
\label{subsec_notation}
To formalize our approach for analyzing these data using a matched design, we introduce some notation.
Consider $I$ matched sets, indexed by $i = 1,..., I$; each with $J_i\textcolor{black}{+1}$ individuals, indexed by $j = 1, ..., J_i + 1$; across $T$ timepoints, indexed by $t = 1, ..., T$.
\textcolor{black}{In our study, $I \in \{8094, 15410\}$ for survivor or family matched sets, respectively, per Subsection \ref{subsec_design}; and $T = (2021 - 2008 + 1) \cdot 12 = 168$, per Section \ref{sec_data}.}
Each set has one exposed and $J_i$ \textcolor{black}{not-yet-exposed} individuals.
\textcolor{black}{For simplicity, we let unit $j = 1$ be exposed and units $j = 2, ..., J_i\textcolor{black}{+1}$ be not-yet-exposed.}
Let $\bm{X}_{tij}$ be a vector of possibly time-varying observed covariates for individual $j$ in set $i$ at time $t$.
Similarly, let $U_{tij}$ represent an unobserved covariate for individual $j$ in set $i$ at time $t$.
Denote by $Z_{ij} \in \left\{\{1, ..., T\} \cup \{\infty\}\right\}$ the time that unit $j$ in set $i$ experiences the exposure (where $Z_{ij} = \infty$ indicates the unit is never exposed). 
Let $Y_{tij}$ be the outcome of interest at time $t$.
For a generic variable $V_{tij}$, denote by $\overline{\bm{V}}_{tij}$ the vector of the history of $V$ up to time $t$ for individual $j$ in set $i$; that is, $\overline{\bm{V}}_{tij} = (V_{1ij}, ..., V_{tij})$.

Following the potential outcomes framework for causal inference \citep{RN28, RN29} each unit has potential outcomes representing the outcome that would be observed had unit $j$ in set $i$ experienced exposure $z \in \left\{\{1, ..., T\} \cup \{\infty\}\right\}$, denoted $y_{tij}(z)$.
Since covariates are time-varying and also possibly affected by the exposure, they are also potential variables, and we similarly denote them using $\bm{x}_{tij}(z)$ and $u_{tij}(z)$ for potential observed and unobserved covariates, respectively.
We make the following assumptions about these potential variables.
\begin{assumption}[Consistency]
\label{eq_sutva}
$Y_{tij} = y_{tij}(Z_{ij})$, $\bm{X}_{tij} = \bm{x}_{tij}(Z_{ij})$, $U_{tij} = u_{tij}(Z_{ij})$.
\end{assumption}
\begin{assumption}[No anticipatory effects]
\label{eq_anticipatory}
$y_{tij}(z) = y_{tij}(\infty)$, $\bm{x}_{tij}(z) = \bm{x}_{tij}(\infty)$, $u_{tij}(z) = u_{tij}(\infty)$ for all $t < z$\textcolor{black}{; and $\bm{x}_{tij}(z) = \bm{x}_{tij}(\infty)$, $u_{tij}(z) = u_{tij}(\infty)$ for all $t = z.$}
\end{assumption}
Assumption \ref{eq_sutva} states that the observed outcome and covariates correspond to the potential outcome and potential covariates under the observed exposure.
Assumption \ref{eq_anticipatory} states that the exposure cannot affect past outcomes or past covariates \textcolor{black}{and, at each timepoint, the covariates are realized before treatment}.
Finally, let
{\color{black}
$$\mathcal{F} := \left\{\left(\bm{x}_{tij}(z),u_{tij}(z)\right): z \in \{\{1, ..., T\} \cup \{\infty\}\}; i = 1, ..., I; j = 1, ..., J_i + 1; t = 1, ..., T \right\}$$
denote the collection of potential covariates} for all units across all timepoints.

\subsection{Estimation and inference for main effects}
\label{subsec_maineff}
\subsubsection{Identification, inference, and sensitivity analysis}
\label{subsubsec_mainident}
For each health outcome, we are interested in estimating the effect either of experiencing a nonfatal a firearm injury or of having a family member experience a nonfatal firearm injury in the month after and the year after the injury.
{\color{black}
Following \cite{Keele2019}, to identify these effects we assume a model of additive bias on the never-treated potential outcomes, which is commonly invoked in difference-in-differences analyses.

\begin{assumption}[Additive bias]
\label{eq_condparallel}
The never-treated potential outcomes are generated as
\begin{align}
\color{black}
\label{eq_addbias}
\left(y_{tij}(\infty) | \mathcal{F}, Z_{ij} = z \right) &= \mu_i + \alpha_{ti} + \beta_{zi} + \epsilon_t\left(\bm{x}_{tij}(z), u_{tij}(z)\right)
\end{align}
for all $i, j, t, z$, where the $\epsilon_t\left(\bm{x}_{tij}(z), u_{tij}(z)\right)$ are independent across matched sets and drawn from a distribution dependent on $\bm{x}_{tij}(z)$, $u_{tij}(z)$, and time period $t$.
\end{assumption}

\textcolor{black}{Let $\epsilon_{tij}(z)$ be shorthand for $\epsilon_t\left(\bm{x}_{tij}(z), u_{tij}(z)\right)$, which encodes the assumption $\epsilon_{tij}(z) \perp Z_{ij} | \mathcal{F}$.}
\textcolor{black}{Assumption \ref{eq_addbias} denotes} the existence of two forms of unobserved additive bias. 
The first, denoted by $\alpha_{ti}$, can vary across matched sets and across time but is constant in exposure status.
\textcolor{black}{The second, denoted by $\beta_{zi}$, can vary across matched sets and exposure status but is constant in time.}
Thus, this assumption aligns with the conditional parallel trends assumption invoked elsewhere in the literature on difference-in-differences with staggered adoption (e.g., \citealt{ DawJamieR2018MaRt, https://doi.org/10.48550/arxiv.2010.04814, RothJonathan2022WTiD}).
That is, Assumption \ref{eq_condparallel} implies,
\begin{align*}
\left(y_{t'ij}(\infty) - y_{tij}(\infty) | Z_{ij} = z\right) &= \alpha_{t'i} - \alpha_{ti} + \left\{\epsilon_{t'ij}\textcolor{black}{(z)} - \epsilon_{tij}\textcolor{black}{(z)}\right\}\\
&= \left(y_{t'ij}(\infty) - y_{tij}(\infty) | Z_{ij} > z\right) 
\end{align*}
for all pairs of time points $t, t' > t$.
The parallel trends assumption may not hold in settings with count or binary outcomes \citep{AtheySusan2006IaIi}, although in our study we have only continuous outcomes.}

\textcolor{black}{We use the difference-in-differences device because, in our setting, there are likely unmeasured characteristics that are time-invariant or vary only very slowly over time (e.g., slower than month-to-month) that affect both the propensity for experiencing a firearm injury (or having a family member injured by firearms) and the health outcomes we analyze.
\textcolor{black}{Model \ref{eq_condparallel} may be violated for a particular outcome if there are unobserved characteristics related both to the outcome and to the probability of firearm injury (or having an injured family member) that vary faster than month-to-month exposure to community violence.}
An alternative identifying assumption is the lagged outcome assumption, where the control potential outcomes are independent of exposure conditional on past outcomes.
However, this assumption is non-nested within the parallel trends assumption and can induce bias if parallel trends instead holds \citep{DawJamieR2018MaRt,ding2019bracketing}.
}

As in \cite{RosenbaumPaulR2020DoOS} and \cite{Keele2019}, we conduct randomization-based inference for treatment effects.
Our test statistic is the outcomes' difference-in-differences across matched sets.
When analyzing outcomes in the month after the injury, this statistic becomes,
\begin{align}
\label{eq_did_month}
    \textcolor{black}{\sum_{i=1}^{I} \sum_{j=1}^{J_i + 1} \left\{I(Z_{ij} = t_i )\left\{Y_{t_iij} - Y_{(t_i-1)ij}\right\} - \frac{I(Z_{ij} > t_i)\left\{Y_{t_iij} - Y_{(t_i - 1)ij}\right\}}{J_i}\right\},}
    \end{align}
where $I(\cdot)$ is the indicator function and $t_i$ is the time that the exposed unit in matched set $i$ was exposed; and for outcomes in the year after the injury, the statistic becomes,
{\color{black}
\begin{align}
\label{eq_did_year}
\sum_{i=1}^{I} \sum_{j=1}^{J_i + 1} &\bigg\{\frac{\sum_{s = 0}^{11} I(Z_{ij} = t_i)\left\{Y_{(t_i+s)ij}  - Y_{(t_i+s - 1)ij}\right\}}{12}\notag \\
&\hspace{1cm}- \frac{\sum_{s = 0}^{11} I(Z_{ij} > t_i)\left\{Y_{(t_i+s)ij} - Y_{(t_i+s - 1)ij}\right\}}{12 J_i}\bigg\}.
\end{align}}

\textcolor{black}{To justify randomization inference, we begin with some notation.}
Let, 
$$\mathcal{F}_\epsilon = \mathcal{F} \cup \left\{\epsilon_{tij}(z): t \in \{1, ..., T\}, i \in \{1, ..., I\}, j \in \{1, ..., J_i + 1\}, z \in \{1, ..., T\} \cup \{\infty\} \right\}.$$
Also, let $\mathcal{Z}$ denote the possible treatment assignments given the matched sets: that is, the set of treatment assignments where one unit in each set is exposed \textcolor{black}{at time $t_i$} and the rest are not-yet-exposed \textcolor{black}{by time $t_i$}.

\textcolor{black}{As is standard in difference-in-differences analyses, we conduct inference by assuming that hidden bias only affects the potential outcomes through the additive form in \eqref{eq_addbias}, and so $\epsilon_t\left(\bm{x}_{tij}(z), u_{tij}(z)\right) = \epsilon_t\left(\bm{x}_{tij}(z)\right)$.
Then, assuming that our matching strategy results in $\bm{x}_{tij}(z) = \bm{x}_{tij'}(z)$ for all $j \neq j'$ and $t \leq z$, the $\{\epsilon_{tij}(z)\}$ are independent and identically distributed \textcolor{black}{for $t \leq z$}.
Further, under the sharp null hypothesis that $y_{tij}(z) = y_{tij}(\infty)$ for all $t, i, j$; the difference-in-differences statistic in \eqref{eq_did_month} for matched set $i$ is equal to,
$$D_i := \sum_{j=1}^{J_i + 1} \left\{I(Z_{ij} = t_i) - J_i^{-1} I(Z_{ij} > t_i)\right\} \underbrace{\left\{\epsilon\left(\bm{x}_{t_i ij}(\infty)\right) - \epsilon\left(\bm{x}_{(t_i - 1)ij}(\infty)\right)\right\}}_{:=\delta_{ij}},$$
and thus the randomization distribution for $\sum_{i=1}^{I} D_i$ is entirely determined by the randomization $\bm{Z} \in \mathcal{Z}$ and the $\delta_{ij}$, which are observed and fixed conditional on $\mathcal{F}_\epsilon$ (a similar argument holds for the difference-in-differences statistic in \eqref{eq_did_year}).
As in \cite{RosenbaumPaulR.2007SAfm}, we can then use the permutational $t$-test for matched sets of varying size with the difference-in-differences test statistics and an approximation to the randomization distribution to calculate $p$-values.}
Confidence intervals are obtained by inverting the hypothesis test.
\textcolor{black}{While our} analysis focuses on testing sharp null hypotheses of no treatment effect, it can be extended to weak nulls following \cite{FogartyColinB.2022Twni}.

In observational studies, it is important to assess the robustness of findings to violations of the key identifying assumptions.
In this paper, we adopt a model for sensitivity analysis in difference-in-differences due to \cite{Keele2019}\textcolor{black}{, and we extend this model to accommodate longitudinal data, staggered treatment adoption, and matched sets of varying size.}
We begin with the following model for treatment assignment,
\begin{align}
\label{eq_sensi0}
\log\left(\dfrac{\mathbb{P}(Z_{ij} = z | \mathcal{F}_{\epsilon})}{\mathbb{P}(Z_{ij} > z| \mathcal{F}_{\epsilon})}\right) = \kappa \left(\bm{x}_{zij}(z)\right) + \gamma u_{zij}(z),
\end{align}
for all $z \in \{1, ..., T\}$ \textcolor{black}{and $0 \leq u_{zij}(z) \leq 1$}.
\textcolor{black}{Following the principles of \citeauthor{RosenbaumObs2002} (\citeyear{RosenbaumObs2002}, Chapter 4) this model expresses the difference in probability of exposure between matched units, marginalized over the unobserved additive biases from Model (1), in terms of another unobserved covariate.
In other words, if we assume that unit $ij$'s probability of exposure at time $z$ depends on observed and unobserved covariates, Equation (4) then models the effect of an unobserved covariate, distinct from those captured by $\alpha_{zi}$ and $\beta_{zi}$ in Equation (1), on the probability of exposure.}

Define $V_{ij} = I(Z_{ij} = t_i)$.
Conditional on $\mathcal{Z}$, this implies the following about the probability of treatment assignment for a particular matched set,
\begin{align*}
\mathbb{P}\left(\bm{V}_i = \bm{v}_i \mid \mathcal{Z}, \mathcal{F}_{\epsilon}\right) = \dfrac{\sum_{j=1}^{J_i + 1} v_{ij} \exp(\gamma u_{t_iij}(t_i))}{\sum_{j=1}^{J_i + 1} \exp(\gamma  u_{t_iij}(t_i))}.
\end{align*}
For \textcolor{black}{$\Gamma = \exp(\gamma) \geq 1$}, \textcolor{black}{this can equivalently be written as},
\begin{align}
\label{eq_sensi}
\color{black}
\dfrac{1}{\Gamma} \leq \dfrac{\mathbb{P}\left(V_{ij} = 1 | \mathcal{Z}, \mathcal{F}_{\epsilon}\right)}{\mathbb{P}\left(V_{ik} = 1 | \mathcal{Z}, \mathcal{F}_{\epsilon}\right)} \leq \Gamma,
\end{align}
\textcolor{black}{for $j \neq k$ \citep{RosenbaumPaulR.2007SAfm}}.
\textcolor{black}{This model allows the use of existing methods for sensitivity analysis \textcolor{black}{as in \cite{RosenbaumPaulR.2018SAFS}}.}

{\color{black}
Equation \ref{eq_sensi} bounds the extent to which unobserved bias, differing from the form presented in Equation \ref{eq_addbias}, can influence the treatment assignment \textcolor{black}{without materially changing the qualitative conclusions of the study}.
\textcolor{black}{A possible interpretation of $\Gamma$ is as follows: a significant effect estimate is robust up to level $\Gamma$ if an unobserved confounder that varies both across time and exposure group can influence the \textcolor{black}{probability} of exposure status by at most a factor $\Gamma$ without substantively changing the result (i.e.,  before the effect fails to become statistically significant).}}

\subsubsection{Estimation with balance at the leaves}
\label{subsec_matching}
Our matched design entails categorical and continuous covariates.
We match exactly on the categorical covariates (prescription drug coverage, insurance plan type, binary sex, census-based statistical area, and primary payer category) and approximately match on the continuous ones (age and risk score).
Because adjusting for covariates measured after the exposure can induce selection bias, we condition only on covariates measured before the exposure, as captured in the subscript $t$ (rather than, e.g., $t'$) in Assumption \ref{eq_condparallel}.
Risk set matching \citep{LiYunfeiPaul2001BRSM}, respects this principle by matching each newly exposed unit at time $t$ with controls who are not yet exposed based on covariate information up to time $t$. 
\textcolor{black}{That is, for each time point $t$ (i.e., month and year), we divide the data into units that are exposed at time $t$ and units that have not yet been exposed by time $t$.
The exposed units are matched to the not-yet-exposed units based on covariates measured up to time $t$ and excluded from matching at later time points.}

\textcolor{black}{
While in our setting, it is never the case that a control unit is later exposed (likely due to the fact that our exposure is relatively rare), in settings where many controls are later exposed, one can modify an analysis based on risk set matching to incorporate this information. 
For example, the exposure at the time of matching can serve as an instrument for actual exposure status in an instrumental variables analysis, as proposed in Section 4.5 of \cite{LiYunfeiPaul2001BRSM}.
}

Our matched design is conducted by profile matching \citep{CohnEricR2022PMft}, where each exposed unit's covariate values are used as the covariate profile around which to approximately balance the matched controls: that is, among the exact matches for the categorical covariates, the algorithm finds, for each exposed unit, the largest subsample of remaining controls whose mean age and mean risk score are no more than 2.5 years and 0.5 units, respectively, different from the exposed unit's.
The algorithm is further modified so that no more than five controls are selected for each exposed unit, as the added value of more controls is limited \citep{MingKewei2000Sgib, HavilandAmelia2007CPSM} and selecting more controls for early-exposed units may diminish the quality of available matches for later-exposed units.
Importantly, this implementation of profile matching has the advantage of achieving balance within matched sets rather than simply across them. 
This is particularly useful in our study, where interest lies not only in estimating aggregate effects but also in uncovering effect modification across subgroups discovered from the data.

The validity of effect estimates for subgroups can be strengthened, in part, by observing covariate balance within these subgroups.
Many matching methods, however, do not guarantee balance for subgroups discovered from the data or, more generally, for arbitrary collections of matched sets (e.g., as in the closed testing procedures of Section \ref{subsec_subgroupeff}).
Our form of profile matching, however, secures balance for any aggregation of matched sets: because we control balance at the matched set-level, any subgroups (which necessarily comprise aggregations of these sets) will achieve a similar level of balance.
In the Supplementary Materials, we provide a  justification for our approach, which ensures balance, for example, at the leaves of a grown classification and regression tree (CART), and compare this to other approaches.

\color{black}
In settings with more covariates, it may not be possible to match exactly for all categorical covariates and match within tight neighborhoods of all continuous ones.
In such cases, subject matter knowledge can guide investigators in choosing which covariates to prioritize for stronger control of bias due to imbalance.
Further, while in our setting small imbalances are critical to control bias for discovered subgroups, in other applications more imbalance may be tolerable.
Finally, we note that while any method of covariate adjustment suffers from this curse of dimensionality, with profile matching, the problems are made more transparent through the (in)feasability of the optimization problem.
\color{black}

\subsection{Estimation and inference for \textcolor{black}{heterogeneity analyses}}
\label{subsec_subgroupeff}

\subsubsection{\textcolor{black}{Refining the exposure and analyzing these effects}}
\label{subsubsec_submax}
While estimates of the average effects of \textcolor{black}{any} firearm injur\textcolor{black}{y} are valuable insofar as they provide a picture of the effects of gun violence \textcolor{black}{in the aggregate}, \textcolor{black}{policymakers may be interested in understanding whether these effects vary for different levels of the exposure.}
Following \cite{SongZirui2022CiHC}, \textcolor{black}{we refine the definition of the exposure and repeat our analyses to understand whether different exposure levels are driving the overall effects: that is, we assess the effects of} (1) \textcolor{black}{an} injury that involved care in an intensive care unit (ICU) care; (2) \textcolor{black}{an} injury that did not require ICU care; (3) \textcolor{black}{an} injury with a documented intent in the claims data, including assault, self-harm, or law enforcement intervention; and (4) \textcolor{black}{an} injury with a documented lack of intent in the claims data (i.e., unintentional).

\textcolor{black}{For this exploratory analysis}, we extend the submax-method of \cite{LeeKwonsang2018Apat}\textcolor{black}{, albeit on groups defined by levels of the exposure rather than groups defined by covariates.}
Like the original submax-method, we compute \textcolor{black}{group}-specific test statistics and test both global null hypotheses and \textcolor{black}{group}-specific null hypotheses by using the joint distribution of the test statistics.
Unlike the original method, however, our null hypotheses assume not a treatment effect of zero, but rather a treatment effect equal to the \textcolor{black}{overall} average treatment effect, as in \cite{LeeKwonsang2021DHEE}.
However, the true \textcolor{black}{overall} average treatment effect, which involves unobserved potential outcomes, is unknown and is instead estimated.
Thus, following \cite{LeeKwonsang2021DHEE}, we evaluate each hypothesis across a grid of possible values for the \textcolor{black}{overall} average treatment effect, where the grid is formed from the $(1 - \alpha_1)$-confidence interval for this effect.
We then perform the hypothesis tests at level $(1-\alpha_2)$ so that $\alpha_1 + \alpha_2 = \alpha$, where $\alpha$ is our desired overall significance level, to preserve the level of the tests.
Mathematical details are included in the Supplementary Materials.

To control the family-wise error rate, we evaluate \textcolor{black}{group}-specific hypotheses via the closed testing procedure of \cite{MARCUSRUTH1976Octp}, which is less conservative than the Bonferroni inequality in sensitivity analysis \citep{RosenbaumPaulR.2009SAfE, FogartyColinB.2016SAfM}.
We use this closed testing procedure to determine the specific \textcolor{black}{exposure levels exhibiting} significantly different \textcolor{black}{effects} from the \textcolor{black}{overall average}.

\subsubsection{Discovering effect modifiers using classification and regression trees}
\label{subsubsec_denovo}
In addition to assessing \textcolor{black}{whether effects vary by pre-specified levels of the exposure}, we also advance methods for discovering effect modifiers from the data.
As an initial analysis, we extend the approach of \cite{DingPeng2019DTEV} from randomized experiments to observational studies.
The method estimates the proportion of treatment effect variation explainable by observed covariates, all under a randomization-based framework.
Outcomes with more explainable variation offer a greater opportunity to discover subgroups with larger-than-average effects.
 
Next, we extend the de novo method of \cite{LeeKwonsang2021DHEE} from matched pairs to matched sets of variable size by replacing the standard nonparametric Wilcoxon's signed rank test with the permutational $t$-test for matched sets.
This method randomly splits the data (we use 25\% for training and 75\% for testing, as in the original paper), uses a classification and regression tree (CART) on the first split to discover subgroups with effects that are different from the average, and implements the extended submax-method as described in Section \ref{subsubsec_submax} for these discovered subgroups in the second split.
The random data splitting preserves statistical validity of the submax-method for discovered subgroups.
We also extend the de novo method by implementing closed testing to test subgroup-specific hypotheses in addition to the global hypothesis of no effect modification by any discovered subgroup, in order to control the family-wise error rate.
In the Supplementary Materials, we evaluate this extension in a simulation study based on \cite{LeeKwonsang2021DHEE}.

\section{Impacts of nonfatal firearm injuries on those injured and their families}
\label{sec_impacts}

\subsection{Evaluating the design of the observational study}
\label{subsec_design}
Table \ref{tab_balance} shows that our matching strategy results in closely balanced covariate \textcolor{black}{distributions} across the treatment groups.
For those injured by firearms, \textcolor{black}{98.3\% were matched to five controls, 0.6\% to four controls, 0.4\% to three controls, 0.4\% to two controls, and 0.3\% to one control}.
For family members of the injured, \textcolor{black}{98.8\% were matched to five controls, 0.4\% were matched to four controls, 0.3\% were matched to three controls, 0.3\% were matched to two controls, and 0.3\% were matched to one control}.

\begin{table}[H]
\singlespacing
\color{black}
\caption{\textcolor{black}{Characteristics of People Injured by Firearms, Family Members of Those Injured, and Their Respective Matched Controls}}
\begin{tabular}{l|rrr|rr}

\label{tab_balance}                & Those      &          &  & Family        &          \\
                                                     & Injured by & Matched  &  & Members of    & Matched  \\
Characteristic                                       & Firearms   & Controls &  & Those Injured & Controls \\ \hline
Age (years)                                          & 34.3       & 34.2     &  & 30.6          & 30.6     \\
Female (\%)                                          & 17.0       & 17.0     &  & 56.7          & 56.7     \\
DxCG risk score                                      & 1.07       & 1.06     &  & 0.88          & 0.87     \\
Prescription drug coverage (\%)                      & 86.6       & 86.7     &  & 86.6          & 86.7     \\
Insurance plan type (\%)                             &            &          &  &               &          \\
\quad Health maintenance organization & 14.1       & 14.1     &  & 15.6          & 15.7     \\
\quad Point of service                & 7.2        & 7.1      &  & 6.7           & 6.7      \\
\quad Preferred provider organization & 57.0       & 57.1     &  & 57.0          & 57.0     \\
\quad Consumer directed health plan   & 10.0       & 10.0     &  & 10.5          & 10.5     \\
\quad High-deductible health plan     & 4.6        & 4.6      &  & 5.2           & 5.2      \\
\quad Other                           & 7.1        & 7.1      &  & 4.9           & 4.9      \\ \hline
Sample size                                          & 8,094      & 40,171   &  & 15,410        & 76,593  
\end{tabular}
\end{table}
\color{black}

By design, the imbalances are controlled by profile matching, which places tight constraints on the imbalances within matched sets and guarantees balance for arbitrary aggregations of matched sets into discovered subgroups.

\subsection{Estimating main effects of nonfatal firearm injuries}
\label{subsec_maineff_results}
\begin{figure}[H]
\caption{Plot of Outcomes for Those Injured of Nonfatal Firearm Injuries, Their Families, and Matched Controls}
\label{fig_outcomes}
\includegraphics[scale=0.5]{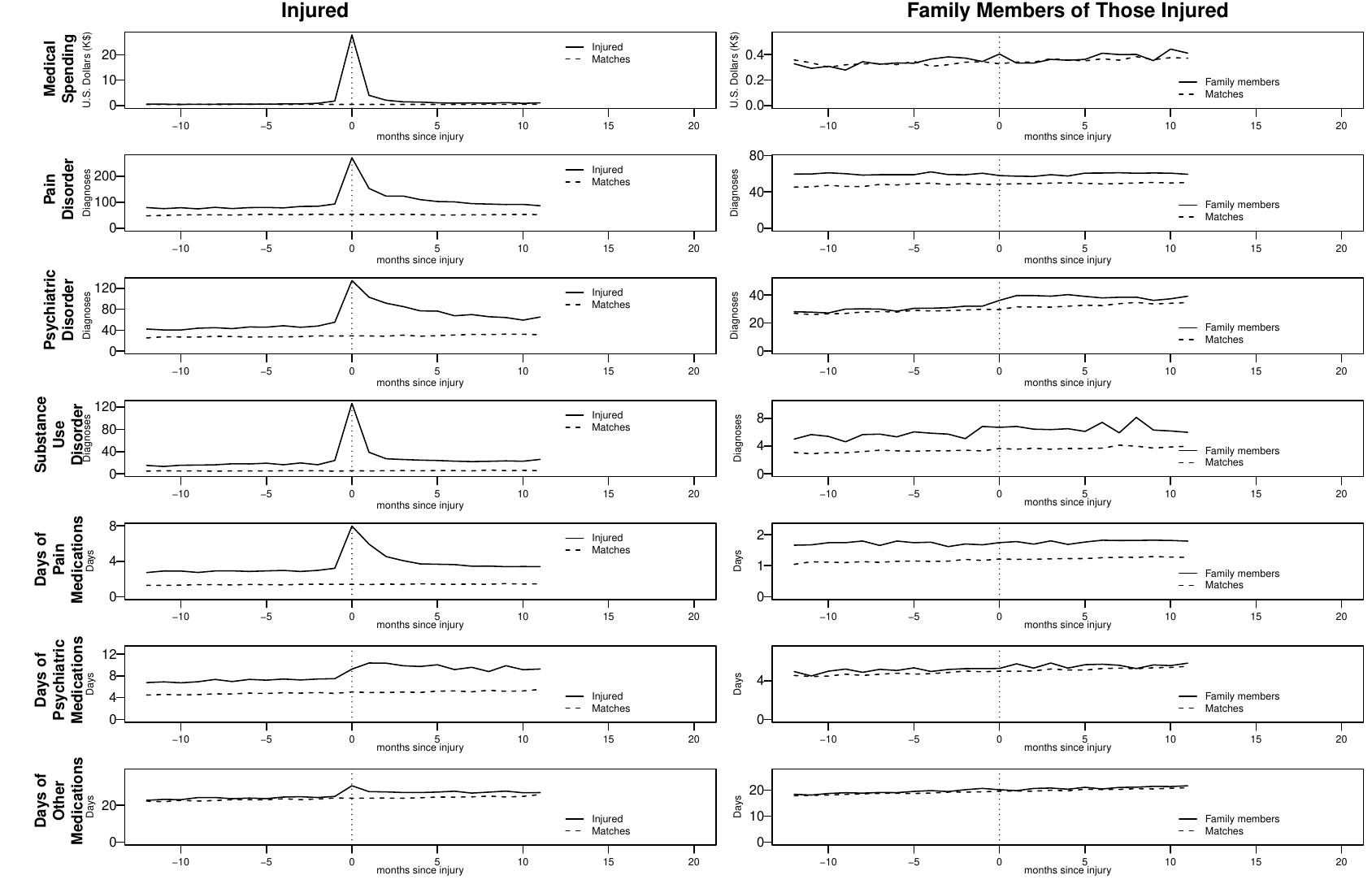}
\end{figure}

As described in Section \ref{sec_data}, we assess the main effects of nonfatal firearm injury on seven outcomes and two timescales for both the injured and their families.
As an initial analysis, Figure \ref{fig_outcomes} plots the outcomes before and after the nonfatal firearm injury, for both the injured and their families, as well as for the matched controls.
\textcolor{black}{The presentation in the figure assumes that effects are homogeneous in time: that is, the figure compares the evolution of outcomes in relative time since the exposure.}
\textcolor{black}{Each timepoint reflects an average across matched sets of contrasts between the exposed unit's and the matched not-yet-exposed units' outcomes, where time is measured relative to the firearm injury.}
The vertical dashed line indicates the time of the injury.
For those injured by firearms, most outcomes exhibit a spike after the injury and no spike for their matched controls, suggesting a large effect of the injury immediately afterward.
In all cases, this increase persists\textcolor{black}{, albeit less pronounced,} throughout the year after injury, particularly for the various diagnoses and medications.
Additionally, the families of those injured may exhibit increases in psychiatric and substance use disorders after their family member's firearm injury.

To formally estimate and test these findings, we use the methods described in Section \ref{subsubsec_mainident}.
Table \ref{tab_main_survivor} presents these results.
In the month immediately afterward,  those injured by firearms experience massive health and financial burdens as a result of their injury.
Medical spending increases by \textcolor{black}{\$27,100}, on average; and pain, psychiatric, and substance use disorder diagnosis rates increase by an average of \textcolor{black}{188.8, 87.9, and 109.2} diagnoses per 1,000 enrollees respectively, on average.
Similar patterns of results are present for the drug prescription outcomes, among those with drug coverage.
Additionally, these effects persist into the year after injury, albeit with smaller magnitude than in the month immediately afterward.
Finally, these results are highly robust to bias due to unmeasured confounding, as indicated by the high values of $\Gamma$ in Table \ref{tab_main_survivor} (see Section \ref{subsubsec_mainident}).
Altogether, the results in Table \ref{tab_main_survivor} demonstrate that nonfatal firearm injuries cause significant health and economic burdens for those injured, and the impacts last for months. 

For family members of those injured by firearms, the effects of their family member's injury are felt most through medical spending and worsened mental health (see Table \ref{tab_main_fam}).
In the month after the injury, family members spend more on health care (\textcolor{black}{\$72}, on average) and receive an average of \textcolor{black}{4.8} more psychiatric disorder diagnoses per 1,000 enrollees.
In the year after injury, the increased medical spending attenuates to non-statistical significance, while the psychiatric disorder diagnosis rates remain high and significant.
These results are sensitive to unmeasured confounding, where an unmeasured confounder that affects the \textcolor{black}{probability} of exposure by \textcolor{black}{1.03 to 1.04} times could substantively change our conclusions.
It is notable also that, while family members experience an increase in psychiatric disorder diagnoses, our data do not reveal a similar increase in prescriptions for these disorders.
\textcolor{black}{This could be consistent with family members receiving other forms of treatment (e.g., psychotherapy) or with an unmet need for medication treatment.}

\begin{sidewaystable}[H]
\singlespacing
\color{black}
\caption{Estimated Effects of Nonfatal Firearm Injury on Those Injured}
\label{tab_main_survivor}
\begin{tabular}{lrrrrrrrrrrr}
                                            & \multicolumn{5}{c}{One Month After Injury}                                                          & \multicolumn{1}{c}{} & \multicolumn{5}{c}{One Year After Injury}                                                           \\ \cline{2-6} \cline{8-12} 
                                            & \multicolumn{1}{c}{} & \multicolumn{1}{c}{} & \multicolumn{2}{c}{Confidence} & \multicolumn{1}{c}{} & \multicolumn{1}{c}{} & \multicolumn{1}{c}{} & \multicolumn{1}{c}{} & \multicolumn{2}{c}{Confidence} & \multicolumn{1}{c}{} \\
                                            & \multicolumn{1}{c}{} & \multicolumn{1}{c}{} & \multicolumn{2}{c}{Interval}   & \multicolumn{1}{c}{} & \multicolumn{1}{c}{} & \multicolumn{1}{c}{} & \multicolumn{1}{c}{} & \multicolumn{2}{c}{Interval}   & \multicolumn{1}{c}{} \\ \cline{4-5} \cline{10-11}
                                            & Point                &                      &                &               &                      &                      & Point                &                      &                &               &                      \\
Outcome                                     & Estimate             & P-value              & Lower          & Upper         & $\Gamma$             &                      & Estimate             & P-value              & Lower          & Upper         & $\Gamma$             \\ \hline
Medical spending (\$)                       & 27100                & $<$0.01              & 26364          & 27829         & $>$10.00             &                      & 2957                 & $<$0.01              & 2855           & 3059          & 3.87                 \\
Diagnosis of…                               &                      &                      &                &               &                      &                      &                      &                      &                &               &                      \\
\quad Pain disorder          & 188.8                & $<$0.01              & 182.7          & 195.0         & 2.24                 &                      & 39.1                 & $<$0.01              & 36.2           & 42.1          & 1.52                 \\
\quad Psychiatric disorder   & 87.9                 & $<$0.01              & 83.5           & 92.2          & 2.00                 &                      & 31.7                 & $<$0.01              & 29.2           & 34.2          & 1.61                 \\
\quad Substance use disorder & 109.2                & $<$0.01              & 105.7          & 112.7         & 4.24                 &                      & 16.0                 & $<$0.01              & 14.8           & 17.3          & 1.92                 \\
Days of prescriptions for…                  &                      &                      &                &               &                      &                      &                      &                      &                &               &                      \\
\quad Pain disorder          & 5.0                  & $<$0.01              & 4.8            & 5.2           & 2.45                 &                      & 1.2                  & $<$0.01              & 1.1            & 1.4           & 1.67                 \\
\quad Psychiatric disorder   & 1.8                  & $<$0.01              & 1.4            & 2.2           & 1.19                 &                      & 2.1                  & $<$0.01              & 1.8            & 2.3           & 1.45                 \\
\quad Other disorders        & 5.8                  & $<$0.01              & 4.6            & 7.0           & 1.20                 &                      & 2.0                  & $<$0.01              & 1.5            & 2.5           & 1.15                
\end{tabular}
\end{sidewaystable}

\begin{sidewaystable}[H]
\singlespacing
\caption{Estimated Effects of Nonfatal Firearm Injury on the Family Members of Those Injured}
\label{tab_main_fam}
\color{black}
\begin{tabular}{lrrrrrrrrrrr}
                                            & \multicolumn{5}{c}{One Month After Injury}                                                          & \multicolumn{1}{c}{} & \multicolumn{5}{c}{One Year After Injury}                                                           \\ \cline{2-6} \cline{8-12} 
                                            & \multicolumn{1}{c}{} & \multicolumn{1}{c}{} & \multicolumn{2}{c}{Confidence} & \multicolumn{1}{c}{} & \multicolumn{1}{c}{} & \multicolumn{1}{c}{} & \multicolumn{1}{c}{} & \multicolumn{2}{c}{Confidence} & \multicolumn{1}{c}{} \\
                                            & \multicolumn{1}{c}{} & \multicolumn{1}{c}{} & \multicolumn{2}{c}{Interval}   & \multicolumn{1}{c}{} & \multicolumn{1}{c}{} & \multicolumn{1}{c}{} & \multicolumn{1}{c}{} & \multicolumn{2}{c}{Interval}   & \multicolumn{1}{c}{} \\ \cline{4-5} \cline{10-11}
                                            & Point                &                      &                &               &                      &                      & Point                &                      &                &               &                      \\
Outcome                                     & Estimate             & P-value              & Lower          & Upper         & $\Gamma$             &                      & Estimate             & P-value              & Lower          & Upper         & $\Gamma$             \\ \hline
Medical spending (\$)                       & 72                   & $<$0.01              & 20             & 124           & 1.03                 &                      & 18                   & 0.21                 & -26            & 62            & NA                   \\
Diagnosis of…                               &                      &                      &                &               &                      &                      &                      &                      &                &               &                      \\
\quad Pain disorder          & -2.8                 & 0.95                 & -6.3           & 0.6           & NA                   &                      & -2.3                 & 0.99                 & -4.2           & -0.4          & NA                   \\
\quad Psychiatric disorder   & 4.8                  & $<$0.01              & 2.2            & 7.4           & 1.04                 &                      & 4.0                  & $<$0.01              & 2.3            & 5.7           & 1.05                 \\
\quad Substance use disorder & 0.7                  & 0.1                  & -0.4           & 1.7           & NA                   &                      & 0.5                  & 0.05                 & -0.1           & 1.0           & NA                   \\
Days of prescriptions for…                  &                      &                      &                &               &                      &                      &                      &                      &                &               &                      \\
\quad Pain disorder          & -0.05                & 0.82                 & -0.17          & 0.06          & NA                   &                      & -0.05                & 0.94                 & -0.11          & 0.01          & NA                   \\
\quad Psychiatric disorder   & -0.09                & 0.72                 & -0.38          & 0.21          & NA                   &                      & 0.0                  & 0.49                 & -0.14          & 0.14          & NA                   \\
\quad Other disorders        & 0.01                 & 0.49                 & -0.73          & 0.75          & NA                   &                      & 0.09                 & 0.29                 & -0.22          & 0.4           & NA                  
\end{tabular}
\end{sidewaystable}

\subsection{Estimating effect modification of nonfatal firearm injuries}
\label{subsec_effmod}

\subsubsection{Analyzing \textcolor{black}{effects for refined exposures}}
\label{subsubsec_submaxres}
We now \textcolor{black}{study the} effects \textcolor{black}{of} the four pre-specified \textcolor{black}{exposure levels} (see Section \ref{subsubsec_submax}).
For those injured by firearms, effects are much larger when the injury is more severe (i.e., requires ICU care) and are moderately larger when the injury is documented in the claims data as a result of an assault, self-harm, or law enforcement intervention.
On the other hand, effects are slightly smaller when the injury is less severe (i.e., does not require ICU care) or when the injury is documented in the claims data as unintentional.
This \textcolor{black}{pattern} is pronounced for medical spending, and in particular among \textcolor{black}{those whose injuries required} ICU \textcolor{black}{care}, for whom medical spending increases by \textcolor{black}{\$121,100} on average the month after injury and by \textcolor{black}{\$13,470} on average the year after injury.

Other notable \textcolor{black}{variation} occurs for the two psychiatric outcomes: those requiring ICU care or whose injury is documented as resulting from an assault, self-harm, or law enforcement \textcolor{black}{intervention} exhibit higher rates of diagnosis and prescription use for psychiatric disorders.
Interestingly, these rates increase throughout the year rather than peaking immediately after the injury, indicating a delayed effect of the exposure for these \textcolor{black}{groups} or delays in obtaining mental health care.
A final pattern emerges among the other prescription outcomes, where those requiring ICU care exhibit higher rates of pain and other medication usage in the month after the injury as compared to other groups of patients, but \textcolor{black}{do not} when considering medication usage over the entire year.
We also assess the robustness of \textcolor{black}{effect estimates} to unobserved confounding, and stronger \textcolor{black}{effect variation} tends to be more robust, as indicated by higher values for $\Gamma$.

Few of \textcolor{black}{the exposure levels exhibit different effects} for families, though there is some evidence \textcolor{black}{of variation} for psychiatric outcomes, medical spending, and substance use disorders; though this \textcolor{black}{is} far less pronounced and more sensitive to unmeasured confounding bias than for the injured.
Relatives of those with more severe injuries exhibit larger-than-average effects for both psychiatric disorder diagnoses and prescription usage.
While this \textcolor{black}{pattern} is present for both the month and the year after their family member's injury for the diagnosis outcome, \textcolor{black}{the effect on} the prescription usage outcome \textcolor{black}{no longer differs from the overall effect} when averaged over the year: a similar pattern as exhibited among the main effects.
Additionally, the effects of a family member's firearm injury on psychiatric disorder diagnoses, in both the month and the year afterward, appear to be largely driven by those whose relative's injury is documented as resulting from self-harm, an assault, or law enforcement intervention.
This suggests an opportunity for targeting mental health interventions for this \textcolor{black}{group}.
Additional results are included in the Supplementary Materials.

\subsubsection{Discovering subgroups with heterogeneous effects}
\label{subsubsec_denovores}
We assess variation in the effects of nonfatal firearm injury for the injured and their families using 9 individual-level and 37 area-level covariates.
Individual covariates include those used for matching as well as the patient's \textcolor{black}{industry of employment}, which is available for those in the \textcolor{black}{MarketScan} database.
Area covariates are measured at the census-based statistical area-level and include aggregate data on educational attainment, race/ethnicity, and household income, as well as measures of health care system availability, utilization, and quality.
A complete list is included in the Supplementary Materials.

The randomization-based $R^2$ bounds for each outcome are presented in Tables \ref{tab_R2_surv} and \ref{tab_R2_fam}.
These can be interpreted as lower and upper bounds on the proportion of variation in the effects of firearm injuries for each outcome that is explained by a linear relationship with the observed covariates (see \citealt{DingPeng2019DTEV} for details).
Despite the quantity and variety of covariates available as potential effect modifiers, they collectively explain relatively low proportions of treatment effect variation across all outcomes and timescales.
For both families and those injured, covariates explain less than 10\% of the variation for all outcomes except for prescriptions for \textcolor{black}{other} disorders, where covariates may explain as much as \textcolor{black}{23\%} of the variation in effects for both those injured and \textcolor{black}{their} families\textcolor{black}{.}

\begin{table}[H]
\singlespacing
\color{black}
\caption{Estimated Variation in Effects of Nonfatal Firearm Injury on Those Injured by Observed Covariates}
\label{tab_R2_surv}
\begin{tabular}{lrrrrr}

                                            & \multicolumn{2}{c}{One Month After Injury}                        & \multicolumn{1}{r}{} & \multicolumn{2}{c}{One Year After Injury}                         \\ \cline{2-3} \cline{5-6} 
                                            & \multicolumn{1}{r}{$R^2$ Lower} & \multicolumn{1}{r}{$R^2$ Upper} & \multicolumn{1}{r}{} & \multicolumn{1}{r}{$R^2$ Lower} & \multicolumn{1}{r}{$R^2$ Upper} \\
                                            & \multicolumn{1}{r}{Bound}       & \multicolumn{1}{r}{Bound}       & \multicolumn{1}{r}{} & \multicolumn{1}{r}{Bound}       & \multicolumn{1}{r}{Bound}       \\ \hline
Medical spending (\$)                       & 0.019                           & 0.022                           &                      & 0.009                           & 0.021                           \\
Diagosis of…                                &                                 &                                 &                      &                                 &                                 \\
\quad Pain disorder          & 0.008                           & 0.021                           &                      & 0.004                           & 0.067                           \\
\quad Psychiatric disorder   & 0.008                           & 0.020                           &                      & 0.006                           & 0.051                           \\
\quad Substance use disorder & 0.008                           & 0.010                           &                      & 0.002                           & 0.007                           \\
Days of prescriptions for…                  &                                 &                                 &                      &                                 &                                 \\
\quad Pain disorder          & 0.007                           & 0.029                           &                      & 0.004                           & 0.032                           \\
\quad Psychiatric disorder   & 0.002                           & 0.046                           &                      & 0.003                           & 0.031                           \\
\quad Other disorders        & 0.002                           & 0.230                           &                      & 0.002                           & 0.095                          
\end{tabular}
\end{table}

\begin{table}[H]
\color{black}
\singlespacing
\caption{Estimated Variation in Effects of Nonfatal Firearm Injury on Family Members of the Injured by Observed Covariates}
\label{tab_R2_fam}
\begin{tabular}{lrrrrr}

                                            & \multicolumn{2}{c}{One Month After Injury} &  & \multicolumn{2}{c}{One Year After Injury} \\ \cline{2-3} \cline{5-6} 
                                            & $R^2$ Lower          & $R^2$ Upper         &  & $R^2$ Lower         & $R^2$ Upper         \\
                                            & Bound                & Bound               &  & Bound               & Bound               \\ \hline
Medical spending (\$)                       & 0.001                & 0.012               &  & 0.000               & 0.000               \\
Diagosis of…                                &                      &                     &  &                     &                     \\
\quad Pain disorder          & 0.001                & 0.048               &  & 0.001               & 0.079               \\
\quad Psychiatric disorder   & 0.001                & 0.035               &  & 0.001               & 0.068               \\
\quad Substance use disorder & 0.001                & 0.004               &  & 0.001               & 0.016               \\
Days of prescriptions for…                  &                      &                     &  &                     &                     \\
\quad Pain disorder          & 0.001                & 0.018               &  & 0.001               & 0.023               \\
\quad Psychiatric disorder   & 0.000                & 0.079               &  & 0.001               & 0.068               \\
\quad Other disorders        & 0.001                & 0.090               &  & 0.001               & 0.159              
\end{tabular}
\end{table}

While relatively few outcomes exhibit large enough explainable variation for subgroup discovery, our approach, which uses CART to discover subgroups with potentially different-from-average effects (see Section \ref{subsubsec_denovo} for details on this method), uncovers a few statistically significant findings.

First, \textcolor{black}{those who are younger (i.e., aged 18 years or less) and live outside of a Metropolitan/Micropolitan Statistical Area show significantly \textcolor{black}{smaller} effects on substance use disorders in the month after injury.
The 95\% confidence interval is (15.1, 36.1).
This interval is much closer to the mean outcome for the control group, which is approximately 5.4, suggesting a much more muted effect on substance use disorders for younger injured individuals.
This result is robust to unmeasured confounding bias ($\Gamma = 2.8$).}

Second, \textcolor{black}{the pain disorder diagnosis outcome exhibits some heterogeneity for survivors.
Those living in areas with few people identifying as multiple races (i.e., percentage in the Metropolitan/Micropolitan Statistical Area indicating two or more races is less than the 25th percentile) show slightly elevated pain diagnoses in the month after the injury.
Further heterogeneity is present for the year after the injury, where those without high risk scores (i.e., risk score falls below the 90th percentile) and live in areas with high rates of uninsurance (i.e., uninsurance rate is above the 80th percentile) show slightly elevated rates of pain disorder diagnoses in the year after their firearm injury.
These results, however, are sensitive to very small unmeasured confounding bias.}

\subsubsection{Re-evaluating the design of the observational study with these subgroups in mind}
\label{subsubsec_design2}
In Section \ref{subsec_design}, we evaluated the design of the observational study by assessing covariate balance between those with nonfatal firearm injuries and their matches, and similarly for the family members of the injured and their matches, as lower covariate imbalance typically corresponds to smaller bias in the average treatment effect estimator \citep{ImaiKosuke2008Mbea}.
We can also assess balance within \textcolor{black}{the discovered subgroups}, which is guaranteed by our profile matching algorithm (see Section \ref{subsec_matching}).

\begin{figure}[H]
\caption{Covariate Balance for \textcolor{black}{the} Discovered Subgroups}
\label{fig_sg_balance}
\includegraphics[scale=0.9]{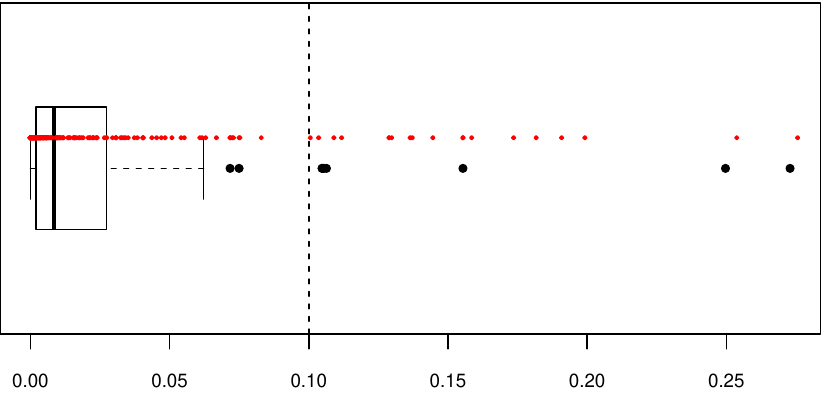}
\flushleft
\footnotesize{Red indicates subgroups that were discovered by CART in the discovery stage but were ultimately rejected in the testing stage.
\textcolor{black}{89.3\% of the comparisons for the statistically significant subgroups and 90.0\% of the comparisons for the non-statistically significant subgroups had ASAMDs $\leq$ 0.1.}}
\end{figure}

Figure \ref{fig_sg_balance} shows the distribution of average standardized absolute mean differences (ASAMDs) for all covariates for \textcolor{black}{the subgroups} discovered from the data.
For a given covariate, the ASAMD is the absolute value of the difference between the exposed and control group means, divided by the covariate's standard deviation.
Heuristically, an ASAMD less than 0.1 is commonly thought to indicate adequate balance, as indicated by the vertical line in the figure.
We see that balance \textcolor{black}{meets this standard} for nearly all covariates across all subgroups.
Over \textcolor{black}{89.3}\% of comparisons \textcolor{black}{for statistically significant subgroups} have ASAMDs less than 0.1, and the few that are greater are from particularly small strata of discovered subgroups \textcolor{black}{and are no larger than 0.28}.
The imbalance resulting from aggregating matched sets into small strata derives from our particular matching method, which does not require sets to be of the same size.
A sensitivity analysis might exclude sets with fewer than five controls (of which there are few; see Section \ref{subsec_design}) to see whether results are similar, as including only sets of the same size will necessarily result in aggregate balance at the level specified from profile matching (see Supplementary Materials).
Additionally, we note that while aggregate covariate balance may be inadequate, there is still balance within each matched set.

\section{Concluding remarks}
\label{sec_conclusion}
In this paper, we estimate the effects of nonfatal firearm injury on those injured and their families using data from \textcolor{black}{a} large health care claims database.
In doing so, we present a framework for estimation, inference, and sensitivity analysis for effect modification in difference-in-differences with staggered exposures, helping to address an important gap in the observational studies literature.
We develop a matching technique, which combines profile matching and risk set matching, for strong control of covariate balance in aggregate and at the leaves of a grown classification or regression tree (i.e., in subgroups discovered from the data).
This technique also preserves the time alignment of the exposure, covariates, and outcomes, avoiding pitfalls of popular TWFE models for difference-in-differences.
Finally, in the broader context of effect modification, we extend two state-of-the-art methods for testing for the presence of effect modification by both pre-specified and discovered subgroups.

Our study reveals significant, large, and persistent effects on the health care spending and the physical, mental, and behavioral health care utilization of those with nonfatal firearm injuries \textcolor{black}{along with} smaller but still persistent effects on the mental health of their families.
For the injured, these effects \textcolor{black}{vary} primarily by the severity of the injury and its documented intent, and the former \textcolor{black}{variation} is particularly robust to bias due to unmeasured confounding.
The mental health effects for families are most experienced by those whose relative's injury is documented as resulting from an assault, self-harm, or law enforcement intervention.

\textcolor{black}{
We also implemented rigorous data-driven methods to discover potential effect modifiers.
While this approach produced a few suggestive results as described in Section \ref{subsubsec_denovores}, overall, across the two exposure groups (i.e., survivors and families), two time scales, and seven outcomes, our analyses in most cases failed to discover meaningful effect modifiers among 41 covariates and their interactions.
This is likely due to the relatively few individual-level covariates available and all covariates' collective inability to explain much of the variation in treatment effects, as presented in Tables \ref{tab_R2_surv} and \ref{tab_R2_fam}.
This highlights the need for higher-quality data collection efforts to further the research community's understanding of this pressing public health issue.}

\bibliographystyle{asa}
\bibliography{Project6_bib}

\section{Supplementary Materials}

\subsection{Details on the Method for Investigating Effect \textcolor{black}{Variation} by Pre-Specified \textcolor{black}{Levels of the Exposure}}
In this section, we provide the mathematical details for the methods for evaluating effect heterogeneity by pre-specified \textcolor{black}{exposure levels} in Section 4.

Our method builds on the submax-method of \cite{LeeKwonsang2018Apat}.
We first slightly modify the notation from Section 2.
Suppose there are $G$ disjoint groups of matched sets $i = 1, ..., I_g$, with $n_{gi}$ individuals in set $i$, $j = 1, ..., n_{gi}$, one treated individual with $Z_{gij} = t$ for some $t$ and  $n_{gi} - 1$ controls with $Z_{gij} = \infty$.
We form matched sets using the observed covariates $\overline{\bm{x}}_t$ but there may be unobserved covariates $\overline{\bm{u}}_t$ as well, which are not controlled.
Recall that, for time-varying covariates $\bm{v}_t$, $\overline{\bm{v}}_t$ indicates the vector of covariates measured before the exposure at time $t$.

For each \textcolor{black}{exposure level group}, we compute a test statistic $T_g = \sum_{i=1}^{I_g} \sum_{j=1}^{n_{gi}} Z_{gij} q_{tt',gij}$ for scores $q_{tt',gij}$ which are functions of the $\delta_{tt', gij}$, $n_{gi}$, and possibly the $\overline{\bm{x}}_{tgij}$.
Under a particular null hypothesis, the distribution of this test statistic, then, depends only on the random treatment assignment (i.e., which of the $n_{gi}$ units in a matched set is treated).
We can also derive bounds on the distribution of the $G$ independent test statistics in the presence of unmeasured confounding (see \cite{LeeKwonsang2018Apat}, Section 2.2).

Following the notation of \cite{LeeKwonsang2018Apat}, our focus lies on $K$ specified comparisons $k = 1,..., K$ involving the $G$ disjoint groups of matched sets.
These comparisons are not necessarily disjoint.
A comparison is a vector $\bm{c}_k = (c_{1k}, ..., c_{Gk})^T$ with $c_{gk} \in \{0, 1\}$, and it determines which of the groups $g \in \{1, ..., G\}$ are aggregated for a given comparison.
For example, if groups $1, ..., G/2$ are matched sets of those with intentional firearm injuries and groups $G/2 + 1, ..., G$ are matched sets of those whose injuries were unintentional, comparison $\bm{c}_2 = (1, ..., 1, 0, ..., 0)^T$ tests a hypothesis among those with intentional injuries, while comparison $\bm{c}_3 = (0, ..., 0, 1, ..., 1)^T$ tests a hypothesis among those with unintentional injuries.
A given comparison is evaluated using test statistic $S_k = \sum_{g = 1}^{G} c_{gk} T_g$.
We test a global hypothesis of no effect modification in any group as well as $K$ hypotheses of effect modification in each of the overlapping subgroups.

To be more precise about our hypotheses, suppose that we know the treatment effect in the population.
That is, we know that $\mathbb{E}[y_{gtij}(t) - y_{gtij}(\infty)] = \tau_t$.
Then our null hypotheses are of the form $H_{0, k}$: $y_{gtij}(t') - y_{gtij}(\infty) = \tau_t$ for all $t \geq t', i, j$ and $g \in \left \{ \bm{c_k} \cdot (1, ..., G)\right\}$.

Let $\bm{C}$ be the $K \times G$ matrix whose $K$ rows are the $\bm{c}_k^T$, $k = 1, ..., K$.
Let $\bm{\theta}_{\Gamma} = \bm{C}\bm{\mu}_{\Gamma}$ and $\bm{\Sigma}_{\Gamma} = \bm{C} \bm{V}_{\Gamma} \bm{C}^T$, where $\bm{\mu}_{\Gamma}$ is vector of maximum expectations of the $T_g$ under sensitivity analysis parameter $\Gamma$ and $\bm{V}_{\Gamma}$ is the diagonal matrix whose elements are the maximum variances of the $T_g$ (see \cite{LeeKwonsang2018Apat} for details).
Letting $\theta_{\gamma k}$ be the $k$th coordinate of $\bm{\theta}_{\Gamma}$ and $\sigma_{\Gamma k}^2$ be the $k$th diagonal element of $\bm{\Sigma}_{\Gamma}$, we define $D_{\Gamma k} = (S_k - \theta_{\Gamma k})/\sigma_{\Gamma k}$ and $\bm{D}_{\Gamma} = (D_{\Gamma 1}, ..., D_{\Gamma K})^T$.
Also, let $\bm{\rho}_\Gamma$ be the $K \times K$ correlation matrix formed by the element $(k, k')$ in $\bm{\Sigma}_{\Gamma}$ by $\sigma_{\Gamma k} \sigma_{\Gamma k'}$.
Under our model for unmeasured confounding, $\bm{D}_{\Gamma}$ converges in distribution to a $N_k(\bm{0}, \bm{\rho}_{\Gamma})$ random variable.
We can then test the global null hypothesis $H_{0}$: $y_{gtij}(t') - y_{gtij}(\infty) = \tau(t, t')$ for all $g, t \geq t', i, j$ by using the maximum deviate $D_{\Gamma \text{max}} = \text{max}_{1 \leq k \leq K} D_{\Gamma k}$, where the $\alpha$ critical value $\kappa_{\Gamma, \alpha}$ solves:
$$1 - \alpha = \mathbb{P}\left(D_{\Gamma \text{max}} < \kappa_{\Gamma, \alpha} \right) = \mathbb{P}
\left(\dfrac{S_k - \theta_{\Gamma k}}{\sigma_{\Gamma k}} < \kappa_{\Gamma, \alpha}, \ k = 1, ..., K \right)$$
We note that, typically, we assume some structure among the population treatment effects, e.g., that they depend only on time since treatment, i.e., on $t - t'$.
This is the case in our setting.

To test the \textcolor{black}{group}-specific null hypotheses, we implement the closed testing procedure as described in Section 4.

In practice, and in our application, the population treatment effects $\tau(t, t')$ are never known, and instead are estimated.
Now, our test statistics are functions of $\tau(t, t')$, so we can denote them, e.g., by $D_{\Gamma \text{max}}(\tau(t, t'))$.
Suppose that we estimate a $(1 - \alpha_1)$-level confidence interval for $\tau(t, t')$, and denote this by $\text{CI}(1 - \alpha_1)(t, t')$, where $\alpha_1 < \alpha$.
Then, instead of using $D_{\Gamma \text{max}}$ as our test statistic, we use the minimum of the maximum deviate as in \cite{LeeKwonsang2021DHEE}:
$$D_{\Gamma \text{minmax}} = \min_{\tau(t, t') \in \text{CI}(1 - \alpha_1)(t, t')} D_{\Gamma \text{max}}(\tau(t, t'))$$
This leads to $D_{\Gamma \text{minmax}} \leq D_{\Gamma \text{max}}(\tau(t, t')^*)$, where $\tau(t, t')^*$ is the true value of $\tau(t, t')$.
We then modify the critical value to be found at $\alpha_2$ such that $\alpha_1 + \alpha_2 = \alpha$.
If $\kappa_{\Gamma, \alpha_2}$ depends on $\tau_t$ (in other cases, e.g., for Wilcoxon's signed rank sum test, it does not), we can take the maximum critical value across $\text{CI}(1 - \alpha_1)_t$ to preserve the overall level of the test.

\subsection{Balance for Arbitrary Aggregations of Matched Sets}
As described in Section 3.3, matching methods for causal inference traditionally focus on aggregate covariate balance or balance within pre-specified subgroup strata.
This is concerning when subgroups are discovered rather than pre-specified, as either the matching cannot be designed to balance within these strata or it must be repeated after subgroup discovery, blurring the separation between the design and analysis stages of observational studies.
On the other hand, with profile matching, balance is evaluated at the level of the matched sets, ensuring balance for any aggregation of sets into subgroups. 
In this section, we discuss formally how profile matching ensures such balance.

First, suppose that there are a fixed number of controls for each treated unit.
That is, suppose each treated unit is matched to $m$ controls such that, in each matched set $j = 1, ..., J$, $k = 1, ..., K$:
$$\left | \sum_{i=1}^{m+1} Z_{ji} B_k(x_{ji}) - \dfrac{1}{m} \sum_{i=1}^{m+1} (1 - Z_{ji}) B_k(x_{ji})\right| \leq \delta_k$$
That is, we balance the means of the $k$ basis functions of the matched controls toward its value for the treated unit.

For any index set $\mathcal{J}^* \subset \{1, ..., J\}$, for all $k$, the imbalance is given by:
\begin{align*}
& \quad \ \hspace{-2pt} \left| \dfrac{\sum_{j \in \mathcal{J}^*} \sum_{i=1}^{m+1} Z_{ji} B_k(x_{ji})}{|\mathcal{J}^*|} - \dfrac{\sum_{j \in \mathcal{J}^*} \sum_{i=1}^{m+1} (1 - Z_{ji}) B_k(x_{ji})}{m |\mathcal{J}^*|}\right| \\
&\leq |\mathcal{J}^*|^{-1} \sum_{j \in \mathcal{J^*}} \left|\sum_{i=1}^{m+1} Z_{ji} B_k(x_{ji}) - \frac{(1 - Z_{ji})}{m}B_k(x_{ji})\right| \\
&\leq  |\mathcal{J}^*|^{-1} \sum_{j \in \mathcal{J}^*} \delta_k \\
&\leq \delta_k
\end{align*}

Thus, however we choose to aggregate across matched sets, we're ensured to get at least as good balance for that aggregation as we have for the individual sets.

Now, suppose that we have a possibly variable number of matched controls per matched set, such that there are $m_j$ matched controls for set $j$.
Then we have, for $j = 1, ..., J$, $k = 1, ..., K$:
$$\left | \sum_{i=1}^{m_j+1} Z_{ji} B_k(x_{ji}) - \dfrac{1}{m_j} \sum_{i=1}^{m_j+1} (1 - Z_{ji}) B_k(x_{ji})\right| \leq \delta_k$$
and for any index set $\mathcal{J}^*$, for all $k$, the imbalance is given by:
\begin{align*}
& \quad \ \hspace{-2pt} \left| \dfrac{\sum_{j \in \mathcal{J}^*} \sum_{i=1}^{m_j+1} Z_{ji} B_k(x_{ji})}{|\mathcal{J}^*|} - \dfrac{\sum_{j \in \mathcal{J}^*} \sum_{i=1}^{m_j+1} (1 - Z_{ji}) B_k(x_{ji})}{ \sum_{j \in \mathcal{J}^*} m_j}\right| \\
&= \left| \sum_{j \in \mathcal{J}^*} \sum_{i=1}^{m_j+1} \dfrac{Z_{ji}}{|\mathcal{J}^*|}B_k(x_{ji}) - \dfrac{m_j}{\sum_{j \in \mathcal{J}^*} m_j} \dfrac{1 - Z_{ji}}{m_j} B_k(x_{ji})\right|\\
&\leq  |\mathcal{J}^*|^{-1}  \sum_{j \in \mathcal{J}^*} \left |\sum_{i=1}^{m_j+1}Z_{ji} B_k(x_{ji}) - \underbrace{\dfrac{m_j |\mathcal{J}^*| }{\sum_{j \in \mathcal{J}^*} m_j}}_{:=w_j} \dfrac{1 - Z_{ji}}{m_j} B_k(x_{ji}) \right |\\
&\leq  |\mathcal{J}^*|^{-1}  \sum_{j \in \mathcal{J}^*} \left |\sum_{i=1}^{m_j+1}Z_{ji} B_k(x_{ji}) - w_j \dfrac{1 - Z_{ji}}{m_j} B_k(x_{ji}) \right |\\
&=  |\mathcal{J}^*|^{-1}  \sum_{j \in \mathcal{J}^*} \left |\sum_{i=1}^{m_j+1}Z_{ji} B_k(x_{ji}) - w_j \dfrac{1 - Z_{ji}}{m_j} B_k(x_{ji}) + \dfrac{1 - Z_{ji}}{m_j} B_k(x_{ji}) -  \dfrac{1 - Z_{ji}}{m_j} B_k(x_{ji})\right |\\
&=  |\mathcal{J}^*|^{-1}  \sum_{j \in \mathcal{J}^*} \left |\sum_{i=1}^{m_j+1}Z_{ji} B_k(x_{ji})  -  \dfrac{1 - Z_{ji}}{m_j} B_k(x_{ji})  + \dfrac{1 - Z_{ji}}{m_j} B_k(x_{ji})(1 - w_j)\right |\\
&\leq  |\mathcal{J}^*|^{-1}  \bigg\{ \sum_{j \in \mathcal{J}^*} \left |\sum_{i=1}^{m_j+1}Z_{ji} B_k(x_{ji}) - \dfrac{1 - Z_{ji}}{m_j} B_k(x_{ji}) \right | + \sum_{j \in \mathcal{J}^*}|1 - w_j| \left |\sum_{i=1}^{m_j+1} \dfrac{1 - Z_{ji}}{m_j} B_k(x_{ji}) \right |\bigg\} \\
&\leq \delta _k+ \sum_{j \in \mathcal{J}^*}|1 - w_j| \left |\sum_{i=1}^{m_j+1} \dfrac{1 - Z_{ji}}{m_j} B_k(x_{ji}) \right |
\end{align*}

Thus, we are not guaranteed our specified level of balance when the matched sets are of varying size.
However, when the index set $\mathcal{J}^*$ includes only sets of the same size, then the weights $w_j$ are equal to 1, and we are guaranteed balance.
Additionally, if the index set $\mathcal{J}^*$ includes mostly sets of the same size, most weights are close to 1, and so departure from the specified level of imbalance is also relatively low.

As a contrast, consider an investigator who has only worried about overall balance, and assume that we have a fixed number of matched controls per treated unit.
That is, for covariates $k = 1, ..., K$:
$$\left|\dfrac{\sum_{j=1}^{J} \sum_{i=1}^{m+1} Z_{ji}B_k(x_{ji})}{J} - \dfrac{\sum_{j=1}^{J} \sum_{i=1}^{m + 1} (1 - Z_{ji})B_k(x_{ji})}{mJ} \right | \leq \delta_k $$
We want to show that for such a method, we can find a partition of the data that achieves the worst-case imbalance.
Denote by $\beta_k^{(z)}$ the mean of covariate transformation $B_k(x)$ in treatment group $Z = z$ and by $\nu_k^{(z)} > 0$ its variance in group $Z = z$.
Then we know that $|\beta_k^{(1)} - \beta_k^{(0)}| \leq \delta_k$.

Suppose we are in the unfavorable situation that all units but one in each treatment group take on the same value.
Call this value $\gamma^{(z)}_k$ in treatment group $Z = z$.
Suppose without loss of generality that the treated unit with the differing value is the first unit in set $j = 1$ and the control unit is the second unit in set $j = 1$.
Then we know that $B_k(x_{11}) = J \beta^{(1)}_k - (J - 1) \gamma^{(1)}_k$ and $B_k(x_{12}) =mJ \beta^{(0)}_k - (mJ - 1)\gamma^{(0)}_k$.
Then:
\begin{align*}
\left | B_k(x_{11})  - B_k(x_{21}) \right| &= \left|J \beta^{(1)}_k - (J - 1) \gamma^{(1)}_k - mJ \beta^{(0)}_k + (mJ - 1)\gamma^{(0)}_k\right| \\
&=\left|J(\beta_k^{(1)} - \beta_k^{(0)}) - (m-1)\beta_k^{(0)}- (J - 1)(\gamma_k^{(1)} - \gamma_k^{(0)})\right|\\
&\leq J \delta_k + (m-1)\left(\beta_k^{(1)} + \delta_k\right)+(J - 1)\left|\gamma_k^{(1)} - \gamma_k^{(0)}\right| \\
&=\delta_k(J + m - 1) + \beta_k^{(1)}(m-1) + (J - 1)\left|\gamma_k^{(1)} - \gamma_k^{(0)}\right|
\end{align*}

Similarly, based on the variance formulas, we can derive:
\begin{align*}
&\quad \hspace{2pt} \left | B_k(x_{11})  - B_k(x_{21}) \right| \\
&= \left|\beta_k^{(1)} + \sqrt{J \nu^{(1)}_k - (J - 1)(\gamma_k^{(1)} - \beta_k^{(1)})^2} - \beta_k^{(0)} - \sqrt{mJ \nu^{(0)}_k - (mJ - 1)(\gamma_k^{(0)} - \beta_k^{(0)})^2}\right| \\
&\geq \left|\left|\beta_k^{(1)} + \sqrt{J \nu^{(1)}_k - (J - 1)(\gamma_k^{(1)} - \beta_k^{(1)})^2}\right| - \left| \beta_k^{(0)} + \sqrt{mJ \nu^{(0)}_k - (mJ - 1)(\gamma_k^{(0)} - \beta_k^{(0)})^2}\right| \right|\\
&\geq \left|\left|\beta_k^{(1)} + \sqrt{J \nu^{(1)}_k - (J - 1)(\gamma_k^{(1)} - \beta_k^{(1)})^2}\right| - \left| \beta_j^{(1)} - \delta_k + \sqrt{mJ \nu^{(0)}_k - (mJ - 1)(\gamma_k^{(0)} - \beta_k^{(1)} + \delta_k)^2}\right| \right| \\
\end{align*}

Thus, we can find a partition of the data (namely, consisting of only these two units) that achieves a nontrivial level of imbalance.
Suppose that $\delta_k = 0.1$, $\beta_k^{(1)} = 0$, $\nu_k^{(1)} = \nu_k^{(0)} = 1$,  $\gamma_k^{(1)} = 50$, $\gamma_k^{(0)} = -50$, $J = 100$ and $m = 5$.
Then the upper bound is approximately 9910 and the lower bound is approximately 617.
Thus, our imbalance is far greater than the controlled level $\delta_k$.

\subsection{Complete Results for\textcolor{black}{Groups Defined by Levels of the Exposure}}
Table \ref{tab_subgroup_survivor} includes results for those with nonfatal firearm injuries, and Table \ref{tab_subgroup_fam} includes results for their families.

\begin{sidewaystable}[H]
\singlespacing
\caption{Estimated Effects of Nonfatal Firearm Injury on Those Injured, by \textcolor{black}{Exposure Level}}
\footnotesize
\label{tab_subgroup_survivor}
\hspace*{-1cm}
\begin{tabular}{lrrrrrrrrrrrrrrr}
                                                             & \multicolumn{1}{c}{} & \multicolumn{1}{c}{} & \multicolumn{1}{c}{} & \multicolumn{1}{c}{} & \multicolumn{1}{c}{} & \multicolumn{1}{c}{} & \multicolumn{1}{c}{} & \multicolumn{1}{c}{} & \multicolumn{3}{c}{Assault,}                          & \multicolumn{1}{c}{} & \multicolumn{1}{c}{} & \multicolumn{1}{c}{} & \multicolumn{1}{c}{} \\
                                                             & \multicolumn{1}{c}{} & \multicolumn{1}{c}{} & \multicolumn{1}{c}{} & \multicolumn{1}{c}{} & \multicolumn{1}{c}{} & \multicolumn{1}{c}{} & \multicolumn{1}{c}{} & \multicolumn{1}{c}{} & \multicolumn{3}{c}{Self-Harm,}                        & \multicolumn{1}{c}{} & \multicolumn{1}{c}{} & \multicolumn{1}{c}{} & \multicolumn{1}{c}{} \\
                                                             & \multicolumn{1}{c}{} & \multicolumn{1}{c}{} & \multicolumn{1}{c}{} & \multicolumn{1}{c}{} & \multicolumn{1}{c}{} & \multicolumn{1}{c}{} & \multicolumn{1}{c}{} & \multicolumn{1}{c}{} & \multicolumn{3}{c}{or Law}                            & \multicolumn{1}{c}{} & \multicolumn{3}{c}{}                                               \\
                                                             & \multicolumn{1}{c}{} & \multicolumn{1}{c}{} & \multicolumn{1}{c}{} & \multicolumn{1}{c}{} & \multicolumn{1}{c}{} & \multicolumn{1}{c}{} & \multicolumn{1}{c}{} & \multicolumn{1}{c}{} & \multicolumn{3}{c}{Enforcement}                       & \multicolumn{1}{c}{} & \multicolumn{3}{c}{Unintentional}                                  \\
                                                             & \multicolumn{3}{c}{ICU Care}                                       & \multicolumn{1}{c}{} & \multicolumn{3}{c}{Non-ICU Care}                                   & \multicolumn{1}{c}{} & \multicolumn{3}{c}{Injury}                            & \multicolumn{1}{c}{} & \multicolumn{3}{c}{Injury}                                         \\ \cline{2-4} \cline{6-8} \cline{10-12} \cline{14-16} 
                                                             & \multicolumn{2}{c}{Confidence}              & \multicolumn{1}{c}{} & \multicolumn{1}{c}{} & \multicolumn{2}{c}{Confidence}              & \multicolumn{1}{c}{} & \multicolumn{1}{c}{} & \multicolumn{2}{c}{Confidence} & \multicolumn{1}{c}{} & \multicolumn{1}{c}{} & \multicolumn{2}{c}{Confidence}              & \multicolumn{1}{c}{} \\
                                                             & \multicolumn{2}{c}{Interval}                & \multicolumn{1}{c}{} & \multicolumn{1}{c}{} & \multicolumn{2}{c}{Interval}                & \multicolumn{1}{c}{} & \multicolumn{1}{c}{} & \multicolumn{2}{c}{Interval}   & \multicolumn{1}{c}{} & \multicolumn{1}{c}{} & \multicolumn{2}{c}{Interval}                & \multicolumn{1}{c}{} \\ \cline{2-3} \cline{6-7} \cline{10-11} \cline{14-15}
                                                             & Lower                & Upper                & $\Gamma$             &                      & Lower                & Upper                & $\Gamma$             &                      & Lower          & Upper         & $\Gamma$             &                      & Lower                & Upper                & $\Gamma$             \\ \hline
Medical spending (\$)                                        &                      &                      &                      &                      &                      &                      &                      &                      &                &               &                      &                      &                      &                      &                      \\
\quad Year after injury                       & 13470                & 14570                & 5.92                 &                      & 1485                 & 1648                 & 2.00                 &                      & 4228           & 4850          & 1.55                 &                      & 2161                 & 2372                 & 1.30                 \\
\quad Month after injury                      & 121100               & 130100               & 8.66                 &                      & 14251                & 15178                & 2.45                 &                      & 39421          & 43888         & 1.79                 &                      & 19974                & 21367                & 1.55                 \\
Diagnosis of…                                                &                      &                      &                      &                      &                      &                      &                      &                      &                &               &                      &                      &                      &                      &                      \\
\quad Pain disorder                           &                      &                      &                      &                      &                      &                      &                      &                      &                &               &                      &                      &                      &                      &                      \\
\quad \quad Year after injury  & 90.1                 & 108.3                & 1.55                 &                      & 28.5                 & 34.7                 & $< 1.01$             &                      & 45.4           & 59.1          & 1.00                 &                      & 28.9                 & 36.1                 & NA                   \\
\quad \quad Month after injury & 204.1                & 242.5                & 1.00                 &                      & 178.1                & 191.0                & NA                   &                      & 185.9          & 215.6         & NA                   &                      & 170.1                & 185.1                & NA                   \\
\quad Psychiatric disorder                    &                      &                      &                      &                      &                      &                      &                      &                      &                &               &                      &                      &                      &                      &                      \\
\quad \quad Year after injury  & 129.0                & 148.5                & 2.00                 &                      & 15.7                 & 20.7                 & 1.14                 &                      & 56.3           & 69.1          & 1.30                 &                      & 17.9                 & 23.8                 & 1.05                 \\
\quad \quad Month after injury & 235.3                & 267.2                & 2.00                 &                      & 62.9                 & 71.8                 & 1.10                 &                      & 140.0          & 162.1         & 1.38                 &                      & 58.6                 & 69.0                 & 1.14                 \\
\quad Substance use disorder                  &                      &                      &                      &                      &                      &                      &                      &                      &                &               &                      &                      &                      &                      &                      \\
\quad \quad Year after injury  & 29.9                 & 39.6                 & 1.34                 &                      & 12.4                 & 14.9                 & NA                   &                      & 22.7           & 30.1          & 1.14                 &                      & 12.1                 & 14.9                 & NA                   \\
\quad \quad Month after injury & 158.4                & 182.9                & 1.41                 &                      & 97.8                 & 105.0                & NA                   &                      & 142.5          & 161.0         & 1.26                 &                      & 87.2                 & 95.3                 & 1.10                 \\
Days of prescriptions for…                                   &                      &                      &                      &                      &                      &                      &                      &                      &                &               &                      &                      &                      &                      &                      \\
\quad Pain disorder                           &                      &                      &                      &                      &                      &                      &                      &                      &                &               &                      &                      &                      &                      &                      \\
\quad \quad Year after injury  & 2.2                  & 3.0                  & 1.30                 &                      & 1.0                  & 1.2                  & NA                   &                      & 1.4            & 1.9           & NA                   &                      & 1.0                  & 1.2                  & NA                   \\
\quad \quad Month after injury & 3.5                  & 4.8                  & NA                   &                      & 4.9                  & 5.4                  & NA                   &                      & 3.9            & 4.9           & NA                   &                      & 4.9                  & 5.4                  & NA                   \\
\quad Psychiatric disorder                    &                      &                      &                      &                      &                      &                      &                      &                      &                &               &                      &                      &                      &                      &                      \\
\quad \quad Year after injury  & 7.9                  & 9.8                  & 1.84                 &                      & 0.9                  & 1.4                  & 1.05                 &                      & 3.9            & 5.1           & 1.22                 &                      & 1.0                  & 1.5                  & $< 1.01$             \\
\quad \quad Month after injury & 1.9                  & 4.8                  & NA                   &                      & 1.1                  & 2.0                  & NA                   &                      & 2.0            & 4.0           & NA                   &                      & 0.9                  & 2.0                  & NA                   \\
\quad Other disorders                         &                      &                      &                      &                      &                      &                      &                      &                      &                &               &                      &                      &                      &                      &                      \\
\quad \quad Year after injury  & 4.8                  & 7.8                  & 1.10                 &                      & 0.9                  & 2.0                  & NA                   &                      & 1.4            & 3.7           & NA                   &                      & 1.2                  & 2.5                  & NA                   \\
\quad \quad Month after injury & -1.5                 & 5.3                  & NA                   &                      & 5.0                  & 7.6                  & NA                   &                      & 2.4            & 7.6           & NA                   &                      & 5.1                  & 8.1                  & NA                  
\end{tabular}
\hspace*{-1cm}
\end{sidewaystable}

\begin{sidewaystable}[H]
\singlespacing
\caption{Estimated Effects of Nonfatal Firearm Injury on the Family Members of Those Injured, by Exposure Level}
\label{tab_subgroup_fam}
\hspace*{-1cm}
\footnotesize
\begin{tabular}{lrrrrrrrrrrrrrrr}
                                                             & \multicolumn{1}{c}{} & \multicolumn{1}{c}{} & \multicolumn{1}{c}{} & \multicolumn{1}{c}{} & \multicolumn{1}{c}{} & \multicolumn{1}{c}{} & \multicolumn{1}{c}{} & \multicolumn{1}{c}{} & \multicolumn{3}{c}{Assault,}                          & \multicolumn{1}{c}{} & \multicolumn{1}{c}{} & \multicolumn{1}{c}{} & \multicolumn{1}{c}{} \\
                                                             & \multicolumn{1}{c}{} & \multicolumn{1}{c}{} & \multicolumn{1}{c}{} & \multicolumn{1}{c}{} & \multicolumn{1}{c}{} & \multicolumn{1}{c}{} & \multicolumn{1}{c}{} & \multicolumn{1}{c}{} & \multicolumn{3}{c}{Self-Harm,}                        & \multicolumn{1}{c}{} & \multicolumn{1}{c}{} & \multicolumn{1}{c}{} & \multicolumn{1}{c}{} \\
                                                             & \multicolumn{1}{c}{} & \multicolumn{1}{c}{} & \multicolumn{1}{c}{} & \multicolumn{1}{c}{} & \multicolumn{1}{c}{} & \multicolumn{1}{c}{} & \multicolumn{1}{c}{} & \multicolumn{1}{c}{} & \multicolumn{3}{c}{or Law}                            & \multicolumn{1}{c}{} & \multicolumn{3}{c}{}                                               \\
                                                             & \multicolumn{1}{c}{} & \multicolumn{1}{c}{} & \multicolumn{1}{c}{} & \multicolumn{1}{c}{} & \multicolumn{1}{c}{} & \multicolumn{1}{c}{} & \multicolumn{1}{c}{} & \multicolumn{1}{c}{} & \multicolumn{3}{c}{Enforcement}                       & \multicolumn{1}{c}{} & \multicolumn{3}{c}{Unintentional}                                  \\
                                                             & \multicolumn{3}{c}{ICU Care}                                       & \multicolumn{1}{c}{} & \multicolumn{3}{c}{Non-ICU Care}                                   & \multicolumn{1}{c}{} & \multicolumn{3}{c}{Injury}                            & \multicolumn{1}{c}{} & \multicolumn{3}{c}{Injury}                                         \\ \cline{2-4} \cline{6-8} \cline{10-12} \cline{14-16} 
                                                             & \multicolumn{2}{c}{Confidence}              & \multicolumn{1}{c}{} & \multicolumn{1}{c}{} & \multicolumn{2}{c}{Confidence}              & \multicolumn{1}{c}{} & \multicolumn{1}{c}{} & \multicolumn{2}{c}{Confidence} & \multicolumn{1}{c}{} & \multicolumn{1}{c}{} & \multicolumn{2}{c}{Confidence}              & \multicolumn{1}{c}{} \\
                                                             & \multicolumn{2}{c}{Interval}                & \multicolumn{1}{c}{} & \multicolumn{1}{c}{} & \multicolumn{2}{c}{Interval}                & \multicolumn{1}{c}{} & \multicolumn{1}{c}{} & \multicolumn{2}{c}{Interval}   & \multicolumn{1}{c}{} & \multicolumn{1}{c}{} & \multicolumn{2}{c}{Interval}                & \multicolumn{1}{c}{} \\ \cline{2-3} \cline{6-7} \cline{10-11} \cline{14-15}
                                                             & Lower                & Upper                & $\Gamma$             &                      & Lower                & Upper                & $\Gamma$             &                      & Lower          & Upper         & $\Gamma$             &                      & Lower                & Upper                & $\Gamma$             \\ \hline
Medical spending (\$)                                        &                      &                      &                      &                      &                      &                      &                      &                      &                &               &                      &                      &                      &                      &                      \\
\quad Year after injury                       & -103                 & 118                  & NA                   &                      & -29                  & 67                   & NA                   &                      & -60            & 109           & NA                   &                      & -20                  & 60                   & NA                   \\
\quad Month after injury                      & -14                  & 239                  & NA                   &                      & 10                   & 123                  & NA                   &                      & -96            & 187           & NA                   &                      & 25                   & 145                  & NA                   \\
Diagnosis of…                                                &                      &                      &                      &                      &                      &                      &                      &                      &                &               &                      &                      &                      &                      &                      \\
\quad Pain disorder                           &                      &                      &                      &                      &                      &                      &                      &                      &                &               &                      &                      &                      &                      &                      \\
\quad \quad Year after injury  & -6.2                 & 5.3                  & NA                   &                      & -4.5                 & -0.5                 & NA                   &                      & -6.4           & 2.3           & NA                   &                      & -4.6                 & 0.1                  & NA                   \\
\quad \quad Month after injury & -12.2                & 8.2                  & NA                   &                      & -6.6                 & 0.7                  & NA                   &                      & -16.6          & -1.0          & NA                   &                      & -6.4                 & 2.2                  & NA                   \\
\quad Psychiatric disorder                    &                      &                      &                      &                      &                      &                      &                      &                      &                &               &                      &                      &                      &                      &                      \\
\quad \quad Year after injury  & 8.7                  & 19.4                 & $< 1.01$             &                      & 1.0                  & 4.5                  & NA                   &                      & 3.7            & 11.6          & NA                   &                      & 0.2                  & 4.3                  & NA                   \\
\quad \quad Month after injury & -2.8                 & 13.3                 & NA                   &                      & 2.0                  & 7.5                  & NA                   &                      & 4.5            & 16.5          & NA                   &                      & -0.2                 & 6.2                  & NA                   \\
\quad Substance use disorder                  &                      &                      &                      &                      &                      &                      &                      &                      &                &               &                      &                      &                      &                      &                      \\
\quad \quad Year after injury  & -1.0                 & 2.8                  & NA                   &                      & -0.2                 & 1.0                   & NA                   &                      & -0.2           & 2.5           & NA                   &                      & -0.8                 & 0.5                  & NA                   \\
\quad \quad Month after injury & -2.3                 & 4.0                  & NA                   &                      & -0.4                 & 1.8                  & NA                   &                      & 0.5            & 5.2           & NA                   &                      & -1.7                 & 0.8                  & NA                   \\
Days of prescriptions for…                                   &                      &                      &                      &                      &                      &                      &                      &                      &                &               &                      &                      &                      &                      &                      \\
\quad Pain disorder                           &                      &                      &                      &                      &                      &                      &                      &                      &                &               &                      &                      &                      &                      &                      \\
\quad \quad Year after injury  & -0.2                 & 0.1                  & NA                   &                      & -0.1                 & 0.0                  & NA                   &                      & -0.3           & 0.0           & NA                   &                      & -0.1                 & 0.1                  & NA                   \\
\quad \quad Month after injury & -0.7                 & -0.1                 & NA                   &                      & -0.1                 & 0.1                  & NA                   &                      & -0.6           & -0.1          & NA                   &                      & -0.1                 & 0.2                  & NA                   \\
\quad Psychiatric disorder                    &                      &                      &                      &                      &                      &                      &                      &                      &                &               &                      &                      &                      &                      &                      \\
\quad \quad Year after injury  & 0                    & 0.9                  & NA                   &                      & -0.2                 & 0.1                  & NA                   &                      & -0.4           & 0.3           & NA                   &                      & -0.2                 & 0.2                  & NA                   \\
\quad \quad Month after injury & -1.2                 & 0.7                  & NA                   &                      & -0.4                 & 0.2                  & NA                   &                      & -0.7           & 0.6           & NA                   &                      & -0.3                 & 0.4                  & NA                   \\
\quad Other disorders                         &                      &                      &                      &                      &                      &                      &                      &                      &                &               &                      &                      &                      &                      &                      \\
\quad \quad Year after injury  & -1.8                 & 0.1                  & NA                   &                      & -0.1                 & 0.5                  & NA                   &                      & -0.8           & 0.7           & NA                   &                      & -0.4                 & 0.4                  & NA                   \\
\quad \quad Month after injury & -4.3                 & 0.3                  & NA                   &                      & -0.5                 & 1.0                  & NA                   &                      & -2.1           & 1.5           & NA                   &                      & -0.8                 & 1.0                  & NA                  
\end{tabular}
\hspace*{-1cm}
\end{sidewaystable}

\subsection{Details and Results for Effect Modifier Discovery}
\subsubsection{Covariates Used for Discovery}
We use the following covariates as inputs to the CART algorithm.
Community characteristics are at the core-based statistical area (CBSA)-level, and are from the American Community Survey \citep{acsdat} and the Area Health Resource File \citep{ahrfdat}. 

\begin{table}[H]
\singlespacing
\begin{tabular}{l|l}
Individual Characteristics                               & Community Characteristics                           \\ \hline
Age (years)                                              & Median age                                          \\
Female (0/1)                                             & Percentage of population with                       \\
DxCG risk score                                          & \quad less than a high school degree \\
Prescription drug coverage (0/1)                         & Percentage of population with                       \\
Insurance plan type (HMO, point of service,              & \quad at most a high school degree   \\
\quad preferred provider organization,    & Percentage of population with                       \\
\quad consumer directed health plan,      & \quad more than a high school degree \\
\quad high-deductible health plan, other) & Median household income                             \\
  Source database         & Percentage of population in poverty                 \\
    & Percentage of population white                      \\
                                & Percentage of population Black                      \\
                                          & Percentage of population American Indian/           \\
                                      & \quad Alaska Native                  \\
                                                         & Percentage of population Asian                      \\
                                                         & Percentage of population Native Hawaiian/           \\
                                                         & \quad Pacific Islander               \\
                                                         & Percentage of population other race                 \\
                                                         & Percentage of population two or more   races        \\
                                                         & Percentage of population Hispanic                   \\
                                                         & Percentage of population currently   working        \\
                                                         & Percentage of population without health   insurance \\
                                                         & Total number of urban households                    \\
                                                         & Total number of rural households                    \\
                                                         & Physicians per 100,000                              \\
                                                         & Community Mental Health Centers per   100,000       \\
                                                         & Federally Qualified Health Centers per   100,000    \\
                                                         & Family doctors per 100,000                          \\
                                                         & Psychiatric doctors per 100,000                     \\
                                                         & Public health doctors per 100,000                   \\
                                                         & Short-term hospitals per 100,000                    \\
                                                         & Long-term hospitals per 100,000                     \\
                                                         & Hospital admissions per 100,000                     \\
                                                         & Hospital beds per 100,000                          
\end{tabular}
\end{table}

\subsubsection{Results}
In this section, we include the CART trees that were identified in by the denovo method (see Section 4.3).
We also list the point estimates for each node and the results of the sensitivity analyses.
Results with $\Gamma =$ NA indicate that the null hypothesis of no effect modification by a particular node was not rejected.
Characteristics with an asterisk (``$*$'') denote that these are characteristics of the Metropolitan/Micropolitan Statistical Area.

We first present results for those who live in a Metropolitan/Micropolitan Statistical Area, for whom characteristics linked by these variables are available.
Then, we present results for those participants who live outside a Metropolitan/Micropolitan Statistical Area.

\singlespacing
\textbf{Individuals with Nonfatal Firearm Injuries, Metropolitan/Micropolitan Statistical Areas}
\begin{figure} [H]
\centering \footnotesize \footnotesize
\caption{Estimated Effects of Nonfatal Firearm Injury on the Injured, Discovered Effect Modifiers of Medical Spending Effects in Metropolitan/Micropolitan Statistical Areas}
\begin{subfigure}[t]{0.475\textwidth}
\centering \footnotesize
\caption{Year After Injury}
\begin{forest}
  forked edges,
  for tree={
    grow'=0,
    draw,
    align=l,
  }
  [none
  ]
\end{forest}
\end{subfigure}
\begin{subfigure}[t]{0.475\textwidth}
\caption{Month After Injury}
\centering \footnotesize
\begin{forest}
  forked edges,
  for tree={
    grow'=0,
    draw,
    align=l,
  }
  [none
  ]
\end{forest}
\end{subfigure}
\end{figure}

\begin{figure} [H]
\centering \footnotesize
\caption{Estimated Effects of Nonfatal Firearm Injury on the Injured, Discovered Effect Modifiers of Pain Disorder Diagnosis Effects in Metropolitan/Micropolitan Statistical Areas}
\begin{subfigure}[t]{0.475\textwidth}
\centering \footnotesize
\caption{Year After Injury}
\hspace{0cm}
\begin{minipage}{\textwidth}
\begin{forest}
  forked edges,
  for tree={
    grow'=0,
    draw,
    align=l,
  }
[
    [{\underline{Risk Score $\geq$ 2.6925} \\ 95\% CI $= (6.9, 37.7)$ \\ $\Gamma = 1.0$}
        [{\underline{Risk Score $\geq$ 2.6925},\\ \underline{Percentage Other Race $\geq$ 7.3*} \\ 95\% CI $= (5.2, 40.7)$ \\ $\Gamma < 1.01$}
        ]
        [{\underline{Risk Score $\geq$ 2.6925}, \\ \underline{Percentage Other Race $\geq$ 7.3*} \\ 95\% CI $= (-33.3, 87.6)$ \\ $\Gamma  < 1.01$}
     ]
    ]
    [{\underline{Risk Score $<$ 2.6925} \\ 95\% CI $= (36.4, 43.2)$ \\ $\Gamma = 1.0$}
        [{\underline{Risk Score $<$ 2.6925},\\ \underline{Percentage Uninsured $<$ 18.1*} \\ 95\% CI $= (37.1, 45.1)$ \\ $\Gamma  < 1.01$}
        ]
        [{\underline{Risk Score $<$ 2.6925},\\ \underline{Percentage Uninsured $\geq$ 18.1*} \\ 95\% CI $=(42.4, 63.0)$ \\ $\Gamma  < 1.01$}
        ]
    ]
]
\end{forest}
\end{minipage}
\end{subfigure}

\vspace{2cm}
\begin{subfigure}[t]{0.475\textwidth}
\caption{Month After Injury}
\centering \footnotesize
\begin{forest}
  forked edges,
  for tree={
    grow'=0,
    draw,
    align=l,
  }
[
    [{\underline{Percentage Two or More Races $\geq$ 2.3*} \\ 95\% CI $ = (166.7, 185.6)$ \\ $\Gamma  < 1.01$}]
    [{\underline{Percentage Two or More Races $<$ 2.3*} \\ 95\% CI $ = (198.6, 225.7)$ \\ $\Gamma  < 1.01 $}]
]
\end{forest}
\end{subfigure}
\end{figure}

\begin{figure} [H]
\centering \footnotesize
\caption{Estimated Effects of Nonfatal Firearm Injury on the Injured, Discovered Effect Modifiers of Psychiatric Disorder Diagnosis Effects in Metropolitan/Micropolitan Statistical Areas}
\begin{subfigure}[t]{0.475\textwidth}
\centering \footnotesize
\caption{Year After Injury}
\begin{forest}
 forked edges,
  for tree={
    grow'=0,
    draw,
    align=l,
  }
  [none
  ]
\end{forest}
\end{subfigure}
\begin{subfigure}[t]{0.475\textwidth}
\caption{Month After Injury}
\centering \footnotesize
\begin{forest}
 forked edges,
  for tree={
    grow'=0,
    draw,
    align=l,
  }
  [none
  ]
\end{forest}
\end{subfigure}
\end{figure}
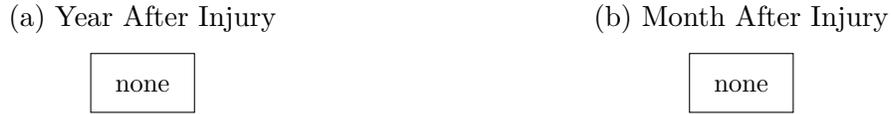

\begin{figure} [H]
\centering \footnotesize
\caption{Estimated Effects of Nonfatal Firearm Injury on the Injured, Discovered Effect Modifiers of Substance Use Disorder Diagnosis Effects in Metropolitan/Micropolitan Statistical Areas}
\begin{subfigure}[t]{0.475\textwidth}
\centering \footnotesize
\caption{Year After Injury}
\begin{forest}
  forked edges,
  for tree={
    grow'=0,
    draw,
    align=l,
  }
  [none
  ]
\end{forest}
\end{subfigure}
\begin{subfigure}[t]{0.475\textwidth}
\caption{Month After Injury}
\centering \footnotesize
\begin{forest}
  forked edges,
  for tree={
    grow'=0,
    draw,
    align=l,
  }
   [none
  ]
\end{forest}
\end{subfigure}
\end{figure}
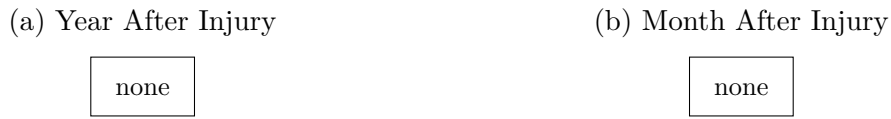

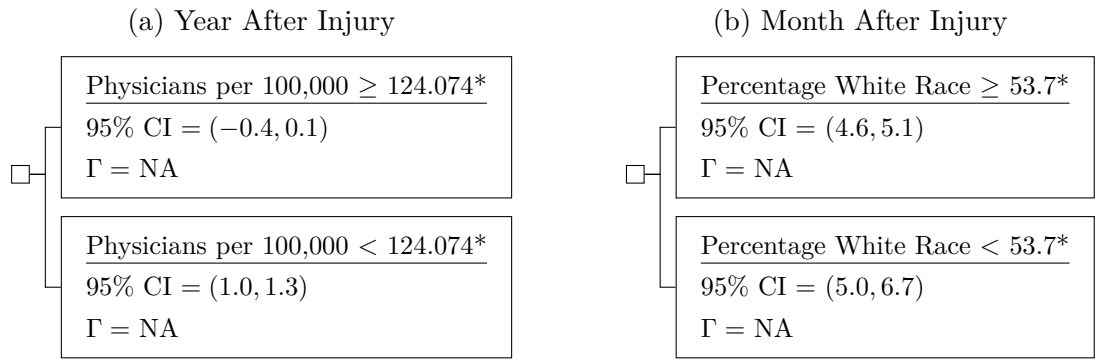
\begin{figure} [H]
\centering \footnotesize
\caption{Estimated Effects of Nonfatal Firearm Injury on the Injured, Discovered Effect Modifiers of Days of Prescriptions for Pain Disorder Effects in Metropolitan/Micropolitan Statistical Areas}
\begin{subfigure}[t]{0.475\textwidth}
\caption{Year After Injury}
\centering \footnotesize
\begin{forest}
 forked edges,
  for tree={
    grow'=0,
    draw,
    align=l,
  }
[
    [{\underline{Physicians per 100,000 $\geq$ 124.074*}\\ 95\% CI $ = (-0.4, 0.1)$ \\ $\Gamma = $ NA}
    ]
    [{\underline{Physicians per 100,000 $<$ 124.074*}\\ 95\% CI $ = (1.0, 1.3)$ \\ $\Gamma = $ NA}
    ]
]
\end{forest}
\end{subfigure}
\begin{subfigure}[t]{0.475\textwidth}
\caption{Month After Injury}
\centering \footnotesize
\begin{forest}
 forked edges,
  for tree={
    grow'=0,
    draw,
    align=l,
  }
  [
    [{\underline{Percentage White Race $\geq$ 53.7*}\\ 95\% CI = $(4.6, 5.1)$\\ $\Gamma = $ NA}
    ]
    [{\underline{Percentage White Race $<$ 53.7*}\\ 95\% CI = $(5.0, 6.7)$\\ $\Gamma = $ NA}
    ]
]
\end{forest}
\end{subfigure}
\end{figure}

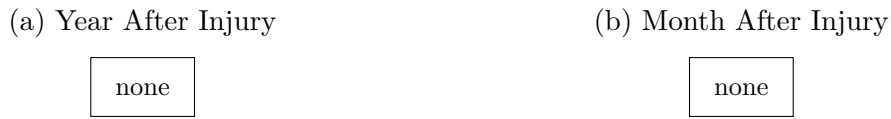
\begin{figure} [H]
\centering \footnotesize
\caption{Estimated Effects of Nonfatal Firearm Injury on the Injured, Discovered Effect Modifiers of Days of Prescriptions for Psychiatric Disorder Effects in Metropolitan/Micropolitan Statistical Areas}
\begin{subfigure}[t]{0.475\textwidth}
\centering \footnotesize
\caption{Year After Injury}
\begin{forest}
  forked edges,
  for tree={
    grow'=0,
    draw,
    align=l,
  }
  [none
  ]
\end{forest}
\end{subfigure}
\begin{subfigure}[t]{0.475\textwidth}
\caption{Month After Injury}
\centering \footnotesize
\begin{forest}
  forked edges,
  for tree={
    grow'=0,
    draw,
    align=l,
  }
   [none
  ]
\end{forest}
\end{subfigure}
\end{figure}

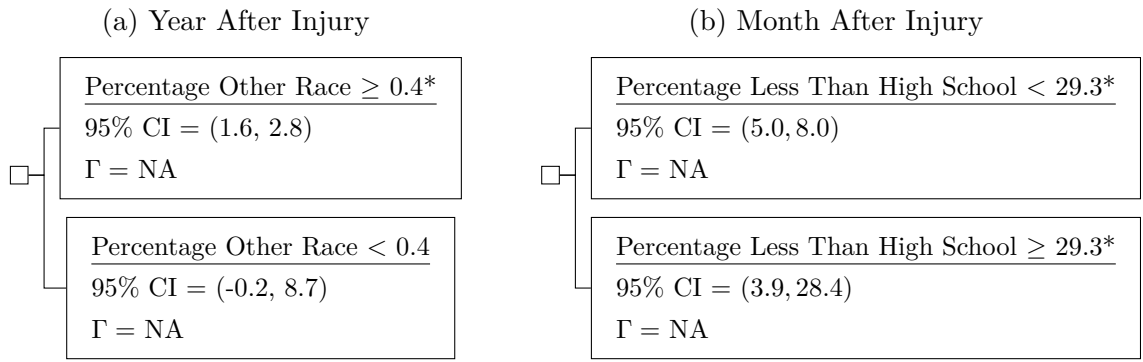
\begin{figure} [H]
\caption{Estimated Effects of Nonfatal Firearm Injury on the Injured, Discovered Effect Modifiers of Days of Prescriptions for Other Disorder Effects in Metropolitan/Micropolitan Statistical Areas}
\begin{subfigure}[t]{0.475\textwidth}
\centering \footnotesize
\caption{Year After Injury}
\begin{forest}
  forked edges,
  for tree={
    grow'=0,
    draw,
    align=l,
  }
 [
	[{\underline{Percentage Other Race $\geq$ 0.4*}\\ 95\% CI\ = (1.6, 2.8)\\ $\Gamma = $ NA}
  ]
	[{\underline{Percentage Other Race $<$ 0.4}\\ 95\% CI\ = (-0.2, 8.7)\\ $\Gamma = $ NA}
	]
 ]
\end{forest}
\end{subfigure}
\begin{subfigure}[t]{0.475\textwidth}
\centering \footnotesize
\caption{Month After Injury}
\begin{forest}
  forked edges,
  for tree={
    grow'=0,
    draw,
    align=l,
  }
 [
	[{\underline{Percentage Less Than High School  $<$ 29.3*}\\ 95\% CI\ = $(5.0, 8.0)$\\ $\Gamma = $ NA}
	]
	[{\underline{Percentage Less Than High School  $\geq$ 29.3*}\\ 95\% CI\ = $(3.9, 28.4)$\\ $\Gamma = $ NA}
	]
 ]
\end{forest}
\end{subfigure}
\end{figure}

\textbf{Family Members of Individuals with Nonfatal Firearm Injuries, Metropolitan/Micropolitan Statistical Areas}
\begin{figure} [H]
\centering \footnotesize
\caption{Estimated Effects of Nonfatal Firearm Injury on Family Members of Those Injured, Discovered Effect Modifiers of Medical Spending Effects in Metropolitan/Micropolitan Statistical Areas}
\begin{subfigure}[t]{0.475\textwidth}
\centering \footnotesize
\caption{Year After Injury}
\begin{forest}
  forked edges,
  for tree={
    grow'=0,
    draw,
    align=l,
  }
  [
  [{\underline{Risk Score $<$ 5.076}\\ 95\% CI = $(-14, 47)$ \\ $\Gamma = $ NA}
  ]
  [{\underline{Risk Score $\geq$ 5.076}\\ 95\% CI = $(-1834, 1694)$ \\ $\Gamma = $ NA}
    [{\underline{Risk Score $\geq$ 5.076},\\ \underline{Percentage Asian Race $<$ 5.8*} \\ 95\% CI = $(-2778, 2604)$ \\ $\Gamma = $ NA}
    ]
    [{\underline{Risk Score $\geq$ 5.076},\\ \underline{Percentage Asian Race $\geq$ 5.8*} \\ 95\% CI = $(-2880, 1459)$ \\ $\Gamma = $ NA}
    ]
  ]
  ]
\end{forest}
\end{subfigure}

\begin{subfigure}[t]{0.475\textwidth}
\caption{Month After Injury}
\hspace{0cm}
\centering \footnotesize
\begin{minipage}{\textwidth}
\begin{forest}
  forked edges,
  for tree={
    grow'=0,
    draw,
    align=l,
  }
  [
  [{\underline{Risk Score $<$ 13.6635}\\ 95\% CI = $(4, 108)$\\ $\Gamma = $ NA}
    [{\underline{Risk Score $<$ 13.6635,}\\ \underline{Percentage White Race $<$ 92.2*}\\ 95\% CI = $(14, 134)$\\ $\Gamma = $ NA}
    ]    
    [{\underline{Risk Score $<$ 13.6635,}\\ \underline{Percentage White Race $\geq$ 92.2*}\\ 95\% CI = $(-583, 359)$ \\ $\Gamma = $ NA}
    ]
  ]
  [{\underline{Risk Score $\geq$ 13.6635}\\95\% CI = $(-9782, 3561 )$ \\ $\Gamma =$ NA}
  ]
  ]
\end{forest}
\end{minipage}
\end{subfigure}
\end{figure}
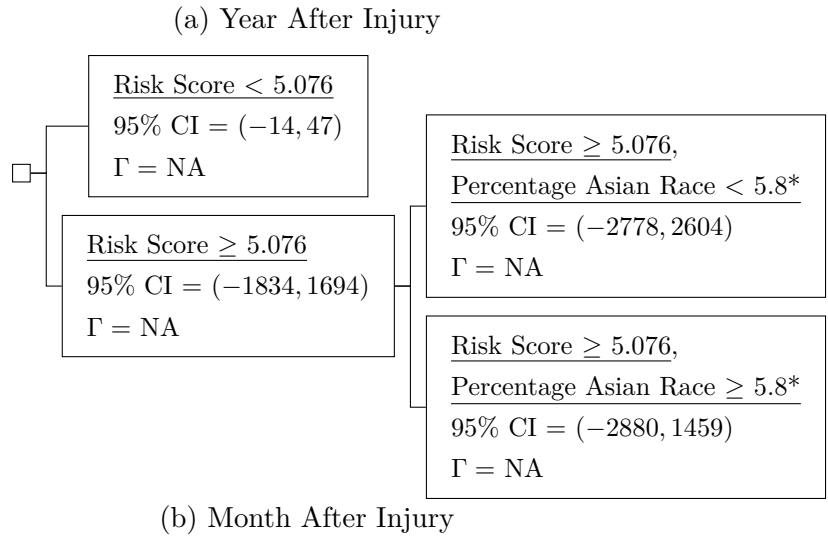

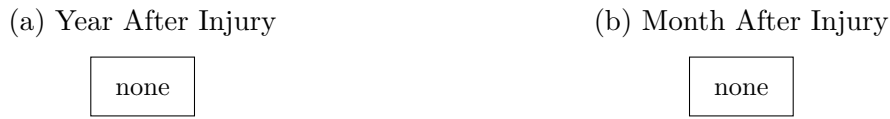
\begin{figure} [H]
\centering \footnotesize
\caption{Estimated Effects of Nonfatal Firearm Injury on Family Members of Those Injured, Discovered Effect Modifiers of Pain Disorder Diagnosis Effects in Metropolitan/Micropolitan Statistical Areas}
\begin{subfigure}[t]{0.475\textwidth}
\centering \footnotesize
\caption{Year After Injury}
\begin{forest}
  forked edges,
  for tree={
    grow'=0,
    draw,
    align=l,
  }
  [none
  ]
\end{forest}
\end{subfigure}
\begin{subfigure}[t]{0.475\textwidth}
\caption{Month After Injury}
\centering \footnotesize
\begin{forest}
  forked edges,
  for tree={
    grow'=0,
    draw,
    align=l,
  }
  [none
  ]
\end{forest}
\end{subfigure}
\end{figure}

\begin{figure} [H]
\centering \footnotesize
\caption{Estimated Effects of Nonfatal Firearm Injury on Family Members of Those Injured, Discovered Effect Modifiers of Psychiatric Disorder Diagnosis Effects in Metropolitan/Micropolitan Statistical Areas}
\begin{subfigure}[t]{0.475\textwidth}
\centering \footnotesize
\caption{Year After Injury}
\begin{forest}
  forked edges,
  for tree={
    grow'=0,
    draw,
    align=l,
  }
  [none
  ]
\end{forest}
\end{subfigure}
\begin{subfigure}[t]{0.475\textwidth}
\caption{Month After Injury}
\centering \footnotesize
\begin{forest}
  forked edges,
  for tree={
    grow'=0,
    draw,
    align=l,
  }
  [none
  ]
\end{forest}
\end{subfigure}
\end{figure}
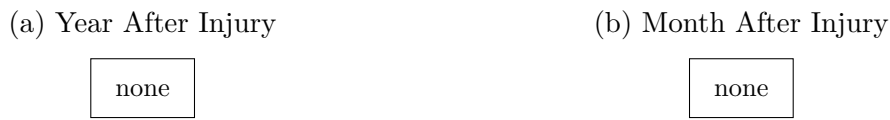

\begin{figure} [H]
\centering \footnotesize
\caption{Estimated Effects of Nonfatal Firearm Injury on Family Members of Those Injured, Discovered Effect Modifiers of Substance Use Disorder Diagnosis Effects in Metropolitan/Micropolitan Statistical Areas}
\begin{subfigure}[t]{0.475\textwidth}
\centering \footnotesize
\caption{Year After Injury}
\begin{forest}
  forked edges,
  for tree={
    grow'=0,
    draw,
    align=l,
  }
 [none
 ]
\end{forest}
\end{subfigure}
\begin{subfigure}[t]{0.475\textwidth}
\centering \footnotesize
\caption{Month After Injury}
\begin{forest}
  forked edges,
  for tree={
    grow'=0,
    draw,
    align=l,
  }
 [none
 ]
\end{forest}
\end{subfigure}
\end{figure}
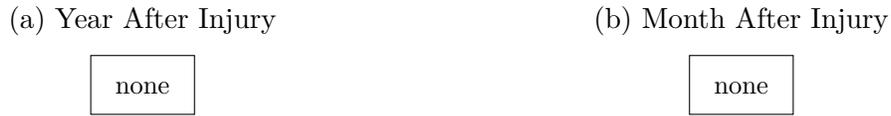

\begin{figure} [H]
\centering \footnotesize
\caption{Estimated Effects of Nonfatal Firearm Injury on Family Members of Those Injured, Discovered Effect Modifiers of Days of Prescriptions for Pain Disorder Effects in Metropolitan/Micropolitan Statistical Areas}
\begin{subfigure}[t]{0.475\textwidth}
\centering \footnotesize
\caption{Year After Injury}
\begin{forest}
  forked edges,
  for tree={
    grow'=0,
    draw,
    align=l,
  }
 [none
 ]
\end{forest}
\end{subfigure}
\begin{subfigure}[t]{0.475\textwidth}
\caption{Month After Injury}
\centering \footnotesize
\begin{forest}
  forked edges,
  for tree={
    grow'=0,
    draw,
    align=l,
  }
 [
    [{\underline{Percentage Working $<$ 71.9*}\\ 95\% CI = $(-0.2, 0.1)$\\ $\Gamma = $ NA}
    ]
    [{\underline{Percentage Working $\geq$ 71.9*}\\ 95\% CI = $(-1.7, 0.2)$\\ $\Gamma = $ NA}
    ]
 ]
\end{forest}
\end{subfigure}
\end{figure}
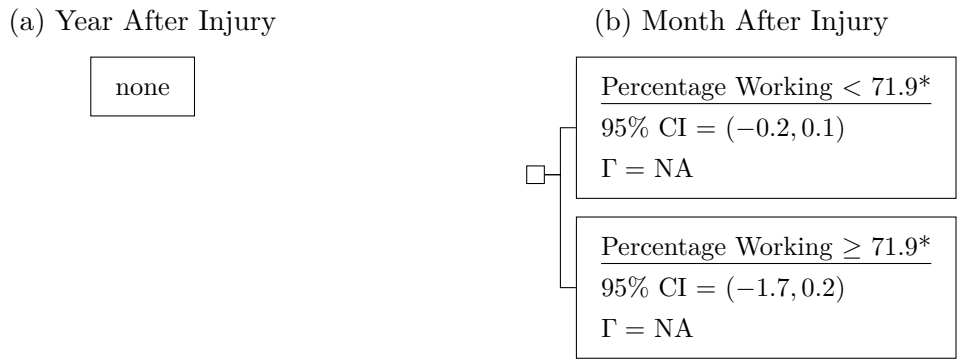

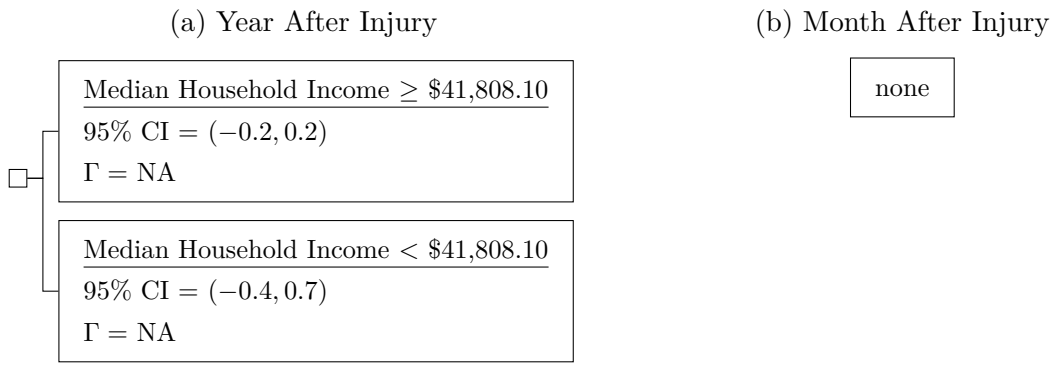
\begin{figure} [H]
\centering \footnotesize
\caption{Estimated Effects of Nonfatal Firearm Injury on Family Members of Those Injured, Discovered Effect Modifiers of Days of Prescriptions for Psychiatric Disorder Effects in Metropolitan/Micropolitan Statistical Areas}
\begin{subfigure}[t]{0.475\textwidth}
\centering \footnotesize
\caption{Year After Injury}
\hspace{0cm}
\begin{minipage}{\textwidth}
\begin{forest}
  forked edges,
  for tree={
    grow'=0,
    draw,
    align=l,
  }
  [
    [{\underline{Median Household Income $\geq$ \$41,808.10}\\ 95\% CI = $(-0.2, 0.2)$ \\ $\Gamma = $ NA}
    ]
    [{\underline{Median Household Income $<$ \$41,808.10}\\ 95\% CI = $(-0.4, 0.7)$ \\ $\Gamma = $ NA}
    ]
  ]
\end{forest}
\end{minipage}
\end{subfigure}
\begin{subfigure}[t]{0.475\textwidth}
\caption{Month After Injury}
\centering \footnotesize
\begin{forest}
  forked edges,
  for tree={
    grow'=0,
    draw,
    align=l,
  }
  [none
  ]
\end{forest}
\end{subfigure}
\end{figure}

\begin{figure} [H]
\centering \footnotesize
\caption{Estimated Effects of Nonfatal Firearm Injury on Family Members of Those Injured, Discovered Effect Modifiers of Days of Prescriptions for Other Disorder Effects in Metropolitan/Micropolitan Statistical Areas}
\begin{subfigure}[t]{0.475\textwidth}
\centering \footnotesize
\caption{Year After Injury}
\begin{forest}
  forked edges,
  for tree={
    grow'=0,
    draw,
    align=l,
  }
  [
 [{\underline{Age $<$ 76.5}\\ 95\% CI = $(-0.3, 0.4)$\\ $\Gamma = $ NA}
 ]
 [{\underline{Age $\geq$ 76.5}\\ 95\% CI = $(-17.8, 15.0)$\\ $\Gamma = $ NA}
 ]
 ]
\end{forest}
\end{subfigure}
\begin{subfigure}[t]{0.475\textwidth}
\caption{Month After Injury}
\centering \footnotesize
\begin{forest}
  forked edges,
  for tree={
    grow'=0,
    draw,
    align=l,
  }
 [none
 ]
\end{forest}
\end{subfigure}
\end{figure}
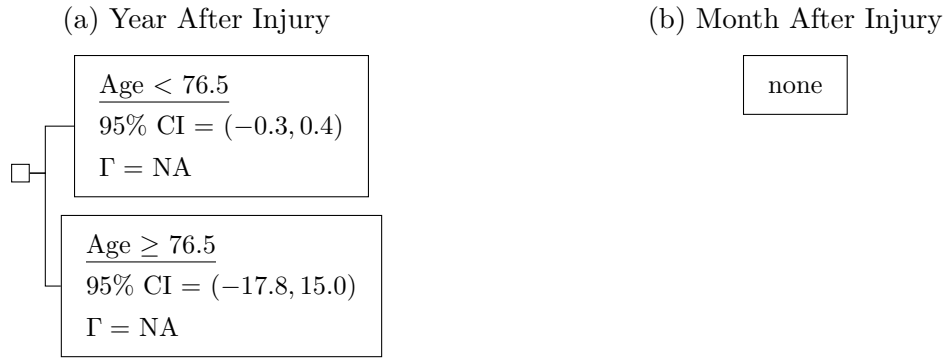

\textbf{Individuals with Nonfatal Firearm Injuries, Non-Metropolitan/Micropolitan Statistical Areas}

\begin{figure} [H]
\centering \footnotesize
\caption{Estimated Effects of Nonfatal Firearm Injury on the Injured, Discovered Effect Modifiers of Medical Spending Effects Outside Metropolitan/Micropolitan Statistical Areas}
\begin{subfigure}[t]{0.475\textwidth}
\centering \footnotesize
\caption{Year After Injury}
\begin{forest}
  forked edges,
  for tree={
    grow'=0,
    draw,
    align=l,
  }
 [none
 ]
\end{forest}
\end{subfigure}
\begin{subfigure}[t]{0.475\textwidth}
\centering \footnotesize
\caption{Month After Injury}
\begin{forest}
  forked edges,
  for tree={
    grow'=0,
    draw,
    align=l,
  }
 [none
 ]
\end{forest}
\end{subfigure}
\end{figure}
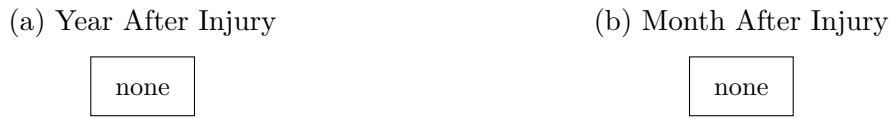

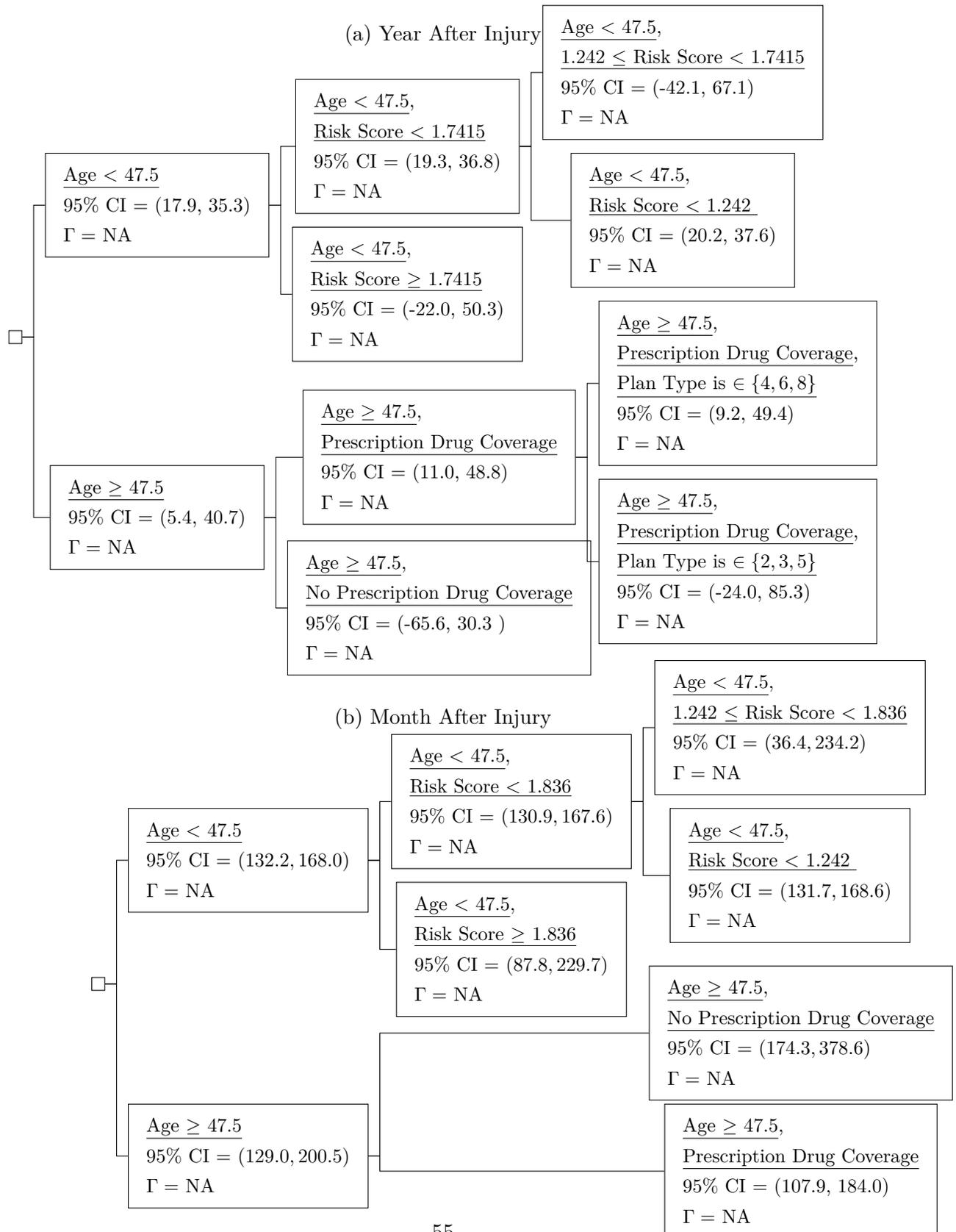
\begin{figure} [H]
\centering \footnotesize
\caption{Estimated Effects of Nonfatal Firearm Injury on the Injured, Discovered Effect Modifiers of Pain Disorder Diagnosis Effects Outside Metropolitan/Micropolitan Statistical Areas}

\begin{subfigure}[t]{0.475\textwidth}
\centering \footnotesize
\caption{Year After Injury}
\vspace{-1cm}
\hspace{-8cm}
\begin{minipage}{\textwidth}
\begin{forest}
  forked edges,
  for tree={
    grow'=0,
    draw,
    align=l,
  }
 [
	[{\underline{Age $<$ 47.5}\\ 95\% CI\ = (17.9, 35.3)\\ $\Gamma = $ NA}
	   [{\underline{Age $<$ 47.5},\\ \underline{Risk Score $<$ 1.7415}\\ 95\% CI\ = (19.3, 36.8)\\ $\Gamma = $ NA}
	       [{\underline{Age $<$ 47.5},\\ \underline{1.242 $\leq$ Risk Score $<$ 1.7415}\\ 95\% CI\ = (-42.1, 67.1)\\ $\Gamma = $ NA}
	       ] 
	       [{\underline{Age $<$ 47.5},\\ \underline{Risk Score $<$ 1.242 }\\ 95\% CI\ = (20.2, 37.6)\\ $\Gamma = $ NA}
	       ] 
	   ] 
     [{\underline{Age $<$ 47.5},\\ \underline{Risk Score $\geq$ 1.7415}\\ 95\% CI\ = (-22.0, 50.3)\\ $\Gamma = $ NA}]
	] 
 [{\underline{Age $\geq$ 47.5}\\ 95\% CI\ = (5.4, 40.7)\\ $\Gamma = $ NA}
    [{\underline{Age $\geq$ 47.5},\\ \underline{Prescription Drug Coverage}\\ 95\% CI\ = (11.0, 48.8)\\ $\Gamma = $ NA}
        [{\underline{Age $\geq$ 47.5},\\ \underline{Prescription Drug Coverage},\\ \underline{Plan Type is $\in \{4, 6, 8\}$}\\ 95\% CI\ = (9.2, 49.4)\\ $\Gamma = $ NA}
        ]
        [{\underline{Age $\geq$ 47.5},\\ \underline{Prescription Drug Coverage},\\ \underline{Plan Type is $\in \{2, 3, 5\}$} \\ 95\% CI\ = (-24.0, 85.3)\\ $\Gamma = $ NA}]
        ]
    [{\underline{Age $\geq$ 47.5},\\ \underline{No Prescription Drug Coverage}\\ 95\% CI\ = (-65.6, 30.3 )\\ $\Gamma = $ NA}]
    ]
 ]
\end{forest}
\end{minipage}
\end{subfigure}

\vspace{0.5cm}
\begin{subfigure}[t]{0.475\textwidth}
\centering \footnotesize
\caption{Month After Injury}
\hspace{-5cm}
\begin{minipage}{\textwidth}
\vspace{-1.5cm}
\begin{forest}
  forked edges,
  for tree={
    grow'=0,
    draw,
    align=l
  }
 [
    [{\underline{Age $<$ 47.5}\\ 95\% CI = $(132.2, 168.0)$ \\ $\Gamma = $ NA}
        [{\underline{Age $<$ 47.5},\\ \underline{Risk Score $<$ 1.836} \\ 95\% CI = $(130.9, 167.6)$ \\ $\Gamma = $ NA}
         [{\underline{Age $<$ 47.5},\\ \underline{1.242 $\leq$ Risk Score $<$ 1.836} \\ 95\% CI = $(36.4, 234.2)$ \\ $\Gamma = $ NA}]
         [{\underline{Age $<$ 47.5},\\ \underline{Risk Score $<$ 1.242 } \\ 95\% CI = $(131.7, 168.6)$ \\ $\Gamma = $ NA}]
        ]
         [{\underline{Age $<$ 47.5},\\ \underline{Risk Score $\geq$ 1.836} \\ 95\% CI = $(87.8, 229.7)$ \\ $\Gamma = $ NA}
         ]
    ]
    [{\underline{Age $\geq$ 47.5}\\ 95\% CI = $(129.0, 200.5)$ \\ $\Gamma = $ NA}
        [{\underline{Age $\geq$ 47.5},\\ \underline{Prescription Drug Coverage}\\ 95\% CI\ = (107.9, 184.0)\\ $\Gamma = $ NA},before computing xy={l=10cm,s=0cm}]
        [{\underline{Age $\geq$ 47.5},\\ \underline{No Prescription Drug Coverage} \\ 95\% CI = $(174.3, 378.6)$ \\ $\Gamma = $ NA},before computing xy={l=10cm,s=2.5cm}
        ]
        ]
    ]
 ]
\end{forest}
\end{minipage}
\end{subfigure}
\end{figure}

\begin{figure} [H]
\centering \footnotesize
\caption{Estimated Effects of Nonfatal Firearm Injury on the Injured, Discovered Effect Modifiers of Psychiatric Disorder Diagnosis Effects Outside Metropolitan/Micropolitan Statistical Areas}
\begin{subfigure}[t]{0.475\textwidth}
\centering \footnotesize
\caption{Year After Injury}
\begin{forest}
  forked edges,
  for tree={
    grow'=0,
    draw,
    align=l,
  }
[none
]
\end{forest}
\end{subfigure}
\begin{subfigure}[t]{0.475\textwidth}
\caption{Month After Injury}
\centering \footnotesize
\begin{forest}
  forked edges,
  for tree={
    grow'=0,
    draw,
    align=l,
  }
[none
]
\end{forest}
\end{subfigure}
\end{figure}
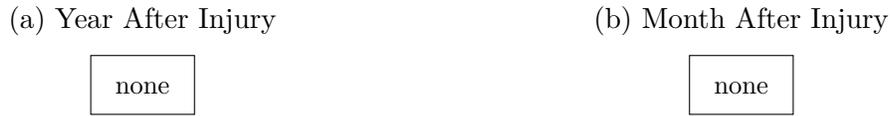

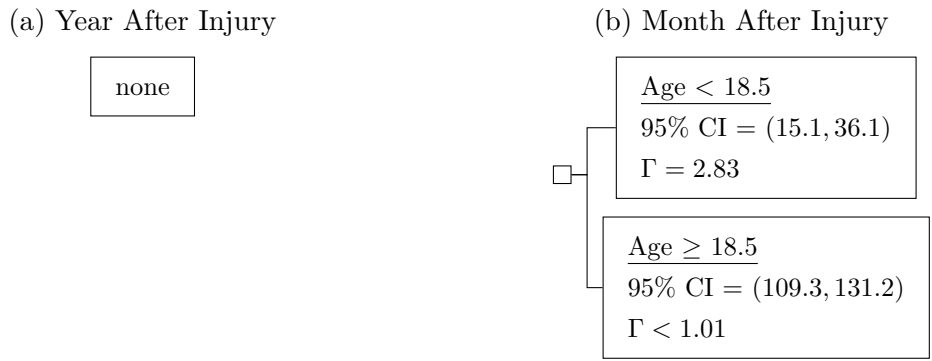
\begin{figure} [H]
\centering \footnotesize
\caption{Estimated Effects of Nonfatal Firearm Injury on the Injured, Discovered Effect Modifiers of Substance Use Disorder Diagnosis Effects Outside Metropolitan/Micropolitan Statistical Areas}
\begin{subfigure}[t]{0.475\textwidth}
\centering \footnotesize
\caption{Year After Injury}
\begin{forest}
  forked edges,
  for tree={
    grow'=0,
    draw,
    align=l,
  }
 [none
 ]
\end{forest}
\end{subfigure}
\begin{subfigure}[t]{0.475\textwidth}
\centering \footnotesize
\caption{Month After Injury}
\begin{forest}
  forked edges,
  for tree={
    grow'=0,
    draw,
    align=l,
  }
 [
    [{\underline{Age $<$ 18.5}\\ 95\% CI = $(15.1, 36.1)$\\ $\Gamma = 2.83$}
    ]
    [{\underline{Age $\geq$ 18.5}\\ 95\% CI = $(109.3, 131.2)$\\ $\Gamma  < 1.01 $}
    ]
 ]
\end{forest}
\end{subfigure}
\end{figure}

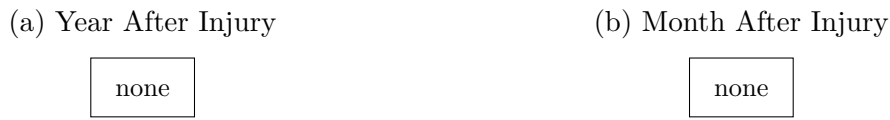
\begin{figure} [H]
\centering \footnotesize
\caption{Estimated Effects of Nonfatal Firearm Injury on the Injured, Discovered Effect Modifiers of Days of Prescriptions for Pain Disorder Effects Outside Metropolitan/Micropolitan Statistical Areas}
\begin{subfigure}[t]{0.475\textwidth}
\centering \footnotesize
\caption{Year After Injury}
\begin{forest}
  forked edges,
  for tree={
    grow'=0,
    draw,
    align=l,
  }
 [none
 ]
\end{forest}
\end{subfigure}
\begin{subfigure}[t]{0.475\textwidth}
\centering \footnotesize
\caption{Month After Injury}
\begin{forest}
  forked edges,
  for tree={
    grow'=0,
    draw,
    align=l,
  }
 [none
 ]
\end{forest}
\end{subfigure}
\end{figure}

\begin{figure} [H]
\centering \footnotesize
\caption{Estimated Effects of Nonfatal Firearm Injury on the Injured, Discovered Effect Modifiers of Days of Prescriptions for Psychiatric Disorder Effects Outside Metropolitan/Micropolitan Statistical Areas}
\begin{subfigure}[t]{0.475\textwidth}
\centering \footnotesize
\caption{Year After Injury}
\begin{forest}
  forked edges,
  for tree={
    grow'=0,
    draw,
    align=l,
  }
 [none
 ]
\end{forest}
\end{subfigure}
\begin{subfigure}[t]{0.475\textwidth}
\centering \footnotesize
\caption{Month After Injury}
\begin{forest}
  forked edges,
  for tree={
    grow'=0,
    draw,
    align=l,
  }
 [none
 ]
\end{forest}
\end{subfigure}
\end{figure}
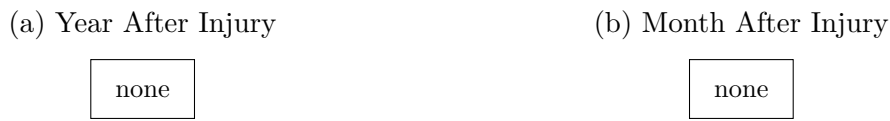

\begin{figure} [H]
\caption{Estimated Effects of Nonfatal Firearm Injury on the Injured, Discovered Effect Modifiers of Days of Prescriptions for Other Disorder Effects Outside Metropolitan/Micropolitan Statistical Areas}
\begin{subfigure}[t]{0.475\textwidth}
\centering \footnotesize
\caption{Year After Injury}
\begin{forest}
  forked edges,
  for tree={
    grow'=0,
    draw,
    align=l,
  }
 [none
 ]
\end{forest}
\end{subfigure}
\begin{subfigure}[t]{0.475\textwidth}
\centering \footnotesize
\caption{Month After Injury}
\begin{forest}
  forked edges,
  for tree={
    grow'=0,
    draw,
    align=l,
  }
 [none
 ]
\end{forest}
\end{subfigure}
\end{figure}
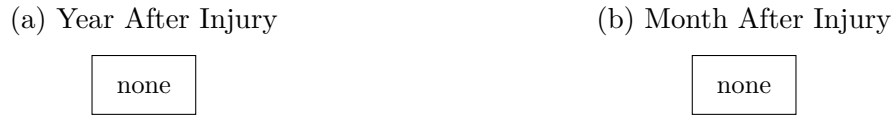

\textbf{Family Members of Individuals with Nonfatal Firearm Injuries, Non-Metropolitan/ Micropolitan Statistical Areas}

\begin{figure} [H]
\centering \footnotesize
\caption{Estimated Effects of Nonfatal Firearm Injury on Family Members of Those Injured, Discovered Effect Modifiers of Medical Spending Effects Outside Metropolitan/Micropolitan Statistical Areas}
\begin{subfigure}[t]{0.475\textwidth}
\centering \footnotesize
\caption{Year After Injury}
\begin{forest}
  forked edges,
  for tree={
    grow'=0,
    draw,
    align=l,
  }
 [none
 ]
\end{forest}
\end{subfigure}
\begin{subfigure}[t]{0.475\textwidth}
\centering \footnotesize
\caption{Month After Injury}
\begin{forest}
  forked edges,
  for tree={
    grow'=0,
    draw,
    align=l,
  }
 [none
 ]
\end{forest}
\end{subfigure}
\end{figure}
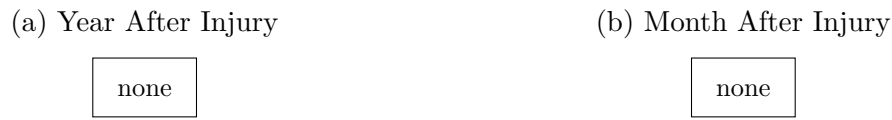

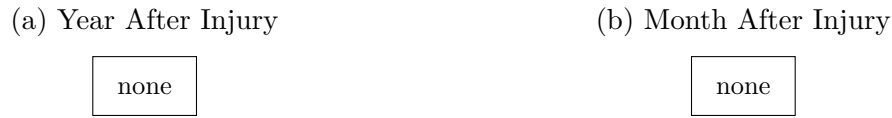
\begin{figure} [H]
\centering \footnotesize
\caption{Estimated Effects of Nonfatal Firearm Injury on Family Members of Those Injured, Discovered Effect Modifiers of Pain Disorder Diagnosis Effects Outside Metropolitan/Micropolitan Statistical Areas}
\begin{subfigure}[t]{0.475\textwidth}
\centering \footnotesize
\caption{Year After Injury}
\begin{forest}
  forked edges,
  for tree={
    grow'=0,
    draw,
    align=l,
  }
 [none
 ]
\end{forest}
\end{subfigure}
\begin{subfigure}[t]{0.475\textwidth}
\centering \footnotesize
\caption{Month After Injury}
\begin{forest}
  forked edges,
  for tree={
    grow'=0,
    draw,
    align=l,
  }
 [none
 ]
\end{forest}
\end{subfigure}
\end{figure}

\begin{figure} [H]
\centering \footnotesize
\caption{Estimated Effects of Nonfatal Firearm Injury on Family Members of Those Injured, Discovered Effect Modifiers of Psychiatric Disorder Diagnosis Effects Outside Metropolitan/Micropolitan Statistical Areas}
\begin{subfigure}[t]{0.475\textwidth}
\centering \footnotesize
\caption{Year After Injury}
\begin{forest}
  forked edges,
  for tree={
    grow'=0,
    draw,
    align=l,
  }
 [none
 ]
\end{forest}
\end{subfigure}
\begin{subfigure}[t]{0.475\textwidth}
\centering \footnotesize
\caption{Month After Injury}
\begin{forest}
  forked edges,
  for tree={
    grow'=0,
    draw,
    align=l,
  }
 [none
 ]
\end{forest}
\end{subfigure}
\end{figure}
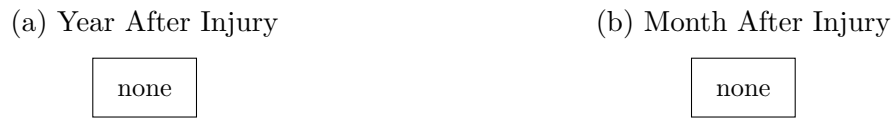

\begin{figure} [H]
\centering \footnotesize
\caption{Estimated Effects of Nonfatal Firearm Injury on Family Members of Those Injured, Discovered Effect Modifiers of Substance Use Disorder Diagnosis Effects Outside Metropolitan/Micropolitan Statistical Areas}
\begin{subfigure}[t]{0.475\textwidth}
\centering \footnotesize
\caption{Year After Injury}
\begin{forest}
  forked edges,
  for tree={
    grow'=0,
    draw,
    align=l,
  }
 [none
 ]
\end{forest}
\end{subfigure}
\begin{subfigure}[t]{0.475\textwidth}
\centering \footnotesize
\caption{Month After Injury}
\begin{forest}
  forked edges,
  for tree={
    grow'=0,
    draw,
    align=l,
  }
 [none
 ]
\end{forest}
\end{subfigure}
\end{figure}
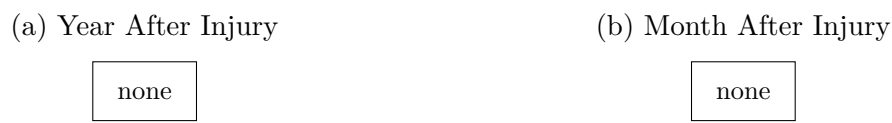

\begin{figure} [H]
\centering \footnotesize
\caption{Estimated Effects of Nonfatal Firearm Injury on Family Members of Those Injured, Discovered Effect Modifiers of Days of Prescriptions for Pain Disorder Effects Outside Metropolitan/Micropolitan Statistical Areas}
\begin{subfigure}[t]{0.475\textwidth}
\centering \footnotesize
\caption{Year After Injury}
\begin{forest}
  forked edges,
  for tree={
    grow'=0,
    draw,
    align=l,
  }
 [none
 ]
\end{forest}
\end{subfigure}
\begin{subfigure}[t]{0.475\textwidth}
\centering \footnotesize
\caption{Month After Injury}
\begin{forest}
  forked edges,
  for tree={
    grow'=0,
    draw,
    align=l,
  }
 [none
 ]
\end{forest}
\end{subfigure}
\end{figure}
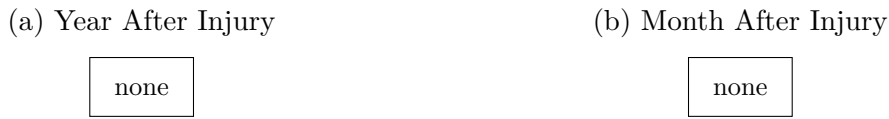

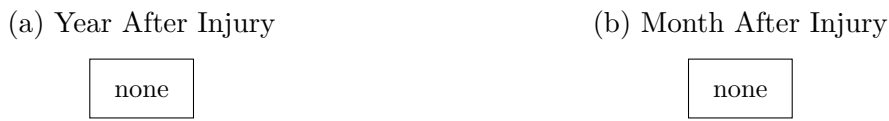
\begin{figure} [H]
\centering \footnotesize
\caption{Estimated Effects of Nonfatal Firearm Injury on Family Members of Those Injured, Discovered Effect Modifiers of Days of Prescriptions for Psychiatric Disorder Effects Outside Metropolitan/Micropolitan Statistical Areas}
\begin{subfigure}[t]{0.475\textwidth}
\centering \footnotesize
\caption{Year After Injury}
\begin{forest}
  forked edges,
  for tree={
    grow'=0,
    draw,
    align=l,
  }
 [none
 ]
\end{forest}
\end{subfigure}
\begin{subfigure}[t]{0.475\textwidth}
\centering \footnotesize
\caption{Month After Injury}
\begin{forest}
  forked edges,
  for tree={
    grow'=0,
    draw,
    align=l,
  }
 [none
 ]
\end{forest}
\end{subfigure}
\end{figure}

\begin{figure} [H]
\caption{Estimated Effects of Nonfatal Firearm Injury on Family Members of Those Injured, Discovered Effect Modifiers of Days of Prescriptions for Other Disorder Effects Outside Metropolitan/Micropolitan Statistical Areas}
\begin{subfigure}[t]{0.475\textwidth}
\centering \footnotesize
\caption{Year After Injury}
\begin{forest}
  forked edges,
  for tree={
    grow'=0,
    draw,
    align=l,
  }
 [none
 ]
\end{forest}
\end{subfigure}
\begin{subfigure}[t]{0.475\textwidth}
\centering \footnotesize
\caption{Month After Injury}
\begin{forest}
  forked edges,
  for tree={
    grow'=0,
    draw,
    align=l,
  }
 [none
 ]
\end{forest}
\end{subfigure}
\end{figure}
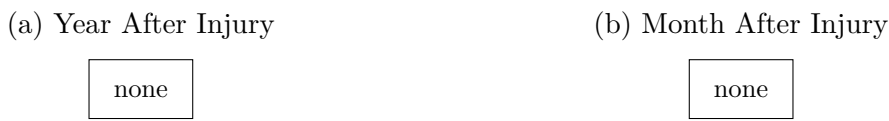

\doublespacing

\subsection{Simulation Study}
We extend the simulation study design from \cite{LeeKwonsang2021DHEE}, where each treated unit is matched to one control, to our setting, where each treated unit is matched to up to five controls.
We begin by briefly describing the simulation study design.

Like \cite{LeeKwonsang2021DHEE}, we vary (1) the choice of tree algorithm, (2) split ratio, and (3) the degree of treatment effect modification.
For (1), we compare the CART and causal tree (CT) \citep{AtheySusan2016Rpfh} methods.
For (2), we consider three ratios for discovery and testing: (10\%, 90\%), (25\%, 75\%), and (50\%, 50\%).
There are $N = 2000$ treated units, and each treated unit is matched to one, two, three, four, or five controls with probabilities $0.025$, $0.025$, $0.05$, $0.1$, and $0.8$, respectively (this gradient reflects the observed gradient in our study).
The true effect size is $0.5$ on average.
There are five covariates, $x_1, ..., x_5$, and at most two of them are true effect modifiers: $x_1$ and/or $x_2$.
For an individual with covariate values $x_1 = a$, $x_2 = b$, we choose a treatment effect from a Normal distribution $N(\tau_{ab}, 1)$.
With $\bm{\tau} = (\tau_{00}, \tau_{01}, \tau_{10}, \tau_{11})$, there are five effect modification settings: (1) $\bm{\tau} = (0.4, 0.4, 0.6, 0.6)$, $\bm{\tau} = (0.3, 0.3, 0.7, 0.7)$, $\bm{\tau} = (0.4, 0.4, 0.5, 0.7)$, $\bm{\tau} = (0.3, 0.3, 0.6, 0.8)$, and $\bm{\tau} = (0.2, 0.5, 0.5, 0.8)$.
We use the permutational $t$-test for matched sets of varying size for testing, and we implement closed testing, as in the main text, to detect which particular effect modifier is driving the main result.

\begin{table}[H]
\singlespacing
\caption{Simulated Power (From 1,000 Replications) for Hypothesis Tests to Discover Heterogeneous Subgroups}
\label{tab_simresults}
\begin{tabular}{llrrrrrrrr}
         Degree of                &    Continuous                                                          & \multicolumn{8}{c}{Splitting ratio (first, second)}                                                                                                  \\ \cline{3-10} 
heterogeneity & outcomes                                          & \multicolumn{2}{c}{(10\%, 90\%)} & \multicolumn{1}{l}{} & \multicolumn{2}{c}{(25\%, 75\%)} & \multicolumn{1}{l}{} & \multicolumn{2}{c}{(50\%, 50\%)} \\ \cline{3-4} \cline{6-7} \cline{9-10} 
($x_1$,   $x_2$)        & $(\tau_{00},   \tau_{01}, \tau_{10}, \tau_{11})$ & CT              & CART           &                      & CT              & CART           &                      & CT              & CART           \\ \hline
(Small, No)             & (0.4, 0.4, 0.6, 0.6)                                         & 0.08            & 0.08           &                      & 0.04            & 0.20           &                      & 0.05            & 0.24           \\
(Large, No)             & (0.3, 0.3, 0.7, 0.7)                                         & 0.48            & 0.90           &                      & 0.58            & 0.90           &                      & 0.86            & 0.90           \\
(Small, Small)          & (0.4, 0.4, 0.5, 0.7)                                         & 0.10            & 0.10           &                      & 0.06            & 0.20           &                      & 0.05            & 0.29           \\
(Large, Small)          & (0.3, 0.3, 0.6, 0.8)                                          & 0.48            & 0.51           &                      & 0.57            & 0.92           &                      & 0.88            & 0.99           \\
(Moderate, Moderate)    & (0.2, 0.5, 0.5, 0.8)                                         & 0.50            & 0.43           &                      & 0.28            & 0.81           &                      & 0.51            & 0.93          
\end{tabular}
\end{table}

Table \ref{tab_simresults} displays the simulated power across the various settings.
When there is low heterogeneity across all variables, both the CART and CT methods have low power for all splitting ratios (though CART shows more power when more data is allowed for the discovery stage).
Larger heterogeneity is more detected for both CT and CART, with CART showing particularly high power across all splitting ratios and CT having highest power for the (50\%, 50\%) splitting ratio.
When two covariates drive moderate effect modification, both methods have moderate power, and CART has high power when more data is allowed for the discovery stage.
Both the CT and CART methods perform the best with the (50\%, 50\%) ratio, though for CART, its performance under (25\%, 75\%) is fairly similar to its performance under (50\%, 50\%).
Additionally, we note that while the magnitude of the results differs from \cite{LeeKwonsang2021DHEE}, the overall pattern is similar.
Differences likely derive from the choice of test statistic, which can impact the power of a test, and/or the differences in the structure of the matches (i.e., matched pairs versus matched sets).

\end{document}